\documentclass[aps,pra,twocolumn,floatfix,nofootinbib,superscriptaddress]{revtex4-2}
\usepackage[colorlinks]{hyperref}
\usepackage[utf8]{inputenc}
\usepackage{graphicx}
\usepackage{tabularx}
\usepackage{multirow}
\usepackage{amssymb}
\usepackage{amsmath}
\usepackage{microtype}
\usepackage{xcolor}
\usepackage{dcolumn}
\usepackage{refcount}

\makeatletter
\newcolumntype{d}[1]{D{.}{.}{#1}}
\makeatother
\def\rmel#1#2#3{\langle #1||#2||#3 \rangle}

\newcommand{\bra}[1]{\ensuremath{\langle #1|}}	
\newcommand{\ket}[1]{\ensuremath{|#1\rangle}}	
\newcommand{\braket}[1]{\ensuremath{\langle #1\rangle}}	

\newcommand{\brad}[1]{\ensuremath{\langle #1||}}
\newcommand{\ketd}[1]{\ensuremath{|| #1\rangle}}
\newcommand{\doublewidetilde}[1]{{\mathpalette\double@widetilde{#1}}}

\def\mr{\mathrm}
\def\mb{\mathbf}

\makeatletter
\begin{document}
    \title{Precision theoretical determination of electric-dipole matrix elements in atomic cesium}
	\author{H. B. Tran Tan}
    \email[Corresponding author: ]{htrantan@unr.edu}
    \affiliation{Department of Physics, University of Nevada, Reno, Nevada 89557, USA}
	\author{A. Derevianko}
	\affiliation{Department of Physics, University of Nevada, Reno, Nevada 89557, USA}
\begin{abstract}
We compute the reduced electric-dipole matrix elements $\brad{nS_{1/2}}D\ketd{n'P_J}$ with $n=6,7$ and $n'=6,7,\ldots,12$ in cesium using the most complete to date \textit{ab initio} relativistic coupled-cluster method which includes singles, doubles, perturbative core triples, and valence triples. Our results agree with previous calculations at the linearized single double level but also show large contributions from nonlinear singles and doubles as well as valence triples. We also calculate the normalized ratio $\xi_{n,n'}\equiv(1/\sqrt{2})\brad{nS_{1/2}}D\ketd{n'P_{1/2}}/\brad{nS_{1/2}}D\ketd{n'P_{3/2}}$ which is important for experimental determination of matrix elements. The ratios $\xi_{6,n}$ display large deviations from the nonrelativistic limit which we associate with Cooper-like minima. Several appendices are provided where we document the procedure for constructing finite basis sets and our implementation of the random phase approximation and Brueckner-orbitals method.
\end{abstract}
\maketitle
\section{Introduction}

Gauging the accuracy of theoretical determinations of atomic parity-violating (APV) amplitudes~\cite{DZUBA1989,Blundell1990,Blundell1992,Porsev2009,Porsev2010,Dzuba2012,SahDasSoi2021-CsPNC,roberts2021comment,BaoDiAndrei2022} generically requires experimental knowledge of three key atomic properties: (i) electric-dipole matrix elements, (ii) magnetic-dipole hyperfine constants, and (iii) atomic energies. While the accuracy of the calculations can be evaluated based on the internal consistency of various many-body approximations and the convergence patterns with respect to the increasing complexity of these approximations, the theory-experiment comparison for known atomic properties remains the key. Indeed, the exact calculations for many-electron atomic systems cannot be carried out in principle, thus leaving the possibility of unaccounted systematic effects. Even for one-electron systems, since the theory is formulated as a perturbation theory in the fine structure constant $\alpha$, the electron-to-nucleus mass ratio, etc., there are always some unaccounted higher-order contributions. Then, only a sufficiently accurate experiment can provide an ``exact'' answer.

In $^{133}\mathrm{Cs}$, where the most accurate to date APV experiment~\cite{Wood1997} has been carried out and new experiments~\cite{Antypas2013,choi2018gain} are planned, the current goal for {\em ab initio} relativistic many-body calculations stands at 0.1\%. As we survey the available experimental data, it becomes clear that the weakest link in the theory-experiment comparison are experimental $E1$ matrix elements (energies are known with high spectroscopic accuracy and the  $^{133}\mathrm{Cs}$ ground state hyperfine splitting is fixed to an exact number by the definition of the unit of time, the second). The accuracy of available experimental $E1$ matrix elements is reviewed below, but it is no better than 0.1\%. Although Ref.~\cite{Toh2019Polarizabilities} reported 0.05\% accuracies for the $\brad{7S_{1/2}}D\ketd{7P_{J}}$ matrix elements, their determination is indirect and relies on theoretical input. More accurate is a direct determination of the ratio of matrix elements. \citet{RafTan98} directly measured the ratio of the cesium ${D}$-line transition strengths, $\left|\left\langle 6 p\,^{2}\!P_{3 / 2}|| D || 6 s\,^{2}\!S_{1 / 2}\right\rangle\right|^{2} /$ $\left|\left\langle 6 p\,^{2} P_{1/2}||D|| 6 s\,^{2}\!S_{1 / 2}\right\rangle\right|^{2}=1.9809(9)$, which translates into a 0.05\%-accurate measurement of the ratio of $E1$ reduced matrix elements. The absorption spectroscopy measurement of the ratio (in contrast to matrix elements) mitigates certain systematic effects, such as dependence on laser power, beam size, collection efficiencies, and detection sensitivities. Another ratio, $\left\langle 7 s\,^{2} \!S_{1 / 2}|| D|| 6 p\,^{2}\!P_{3 / 2}\right\rangle /\left\langle 7 s\,^{2} S_{1 / 2}||D|| 6 p\,^{2}\!P_{1 / 2}\right\rangle= 1.5272(17)$, was recently measured using a two-color, two-photon excitation technique~\cite{Toh2019Mels}.  

Motivated by these experimental developments, the primary goal of this paper is to examine the behavior of the {\em normalized ratio of reduced dipole matrix elements} connecting the initial $nS_{1/2}$ state to the two fine-structure components $n'P_J$,
\begin{equation}\label{eq:effratio}
    \xi_{n,n'}\equiv\frac{1}{\sqrt{2}}\frac{\bra{nS_{1/2}}|D|\ket{n'P_{3/2}}}{\bra{nS_{1/2}}|D|\ket{n'P_{1/2}}}\,.
\end{equation}

We will examine the behavior of the ratio $\xi_{n,n'}$ as a function of the final state principle quantum number $n'$ while fixing the initial state. 
We have chosen the renormalization factor of $1/\sqrt{2}$ so that in
the nonrelativistic limit $\xi_{n,n'} \rightarrow 1$.
Generically, one would expect the relativistic correction to be $\sim(\alpha Z)^2$, which evaluates to 0.16 for Cs. 
However, we find that the normalized ratio~(\ref{eq:effratio}) can substantially deviate from 1, signaling a complete break-down of such an expectation. Moreover, we find that the ratio~(\ref{eq:effratio}) substantially depends on many-body effects included in the calculations. Such significant deviations are known in photoionization processes for alkali-metal atoms~(see, e.g., Ref.~\cite{Derevianko2000_photo} and references therein), where the ground state $n S_{1/2}$ can be ionized into either  the $P_{1/2}$ or $P_{3/2}$ channel. Due to a phase shift between the $\varepsilon P_{3/2}$ and $\varepsilon P_{1/2}$ continuum wave functions of the outgoing electron with energy $\varepsilon$, a situation may arise that the $n S_{1/2}\rightarrow \varepsilon P_{1/2}$ transition amplitude vanishes, while the $n S_{1/2}\rightarrow \varepsilon P_{3/2}$ amplitude does not. In this case, the ratio~\eqref{eq:effratio} based on discrete-to-continuum matrix elements becomes infinite. This is the origin of the Cooper minimum~\cite{FANO1968,Manson1985} in photoionization cross-sections. While in our case of bound-bound transitions, the Cooper minimum does not occur \textit{per se}, similar logic applies to explaining large deviations of  $\xi_{n,n'}$ from 1.

To explore the sensitivity of the ratio~\eqref{eq:effratio} to correlation corrections, we use a variety of relativistic many-body methods ranging from the random-phase Approximation (RPA) and Brueckner-orbital (BO) methods to the more sophisticated coupled-cluster (CC) techniques.  

The secondary goal of this paper is to compile electric-dipole ($E1$) matrix elements $\brad{nS_{1/2}}D\ketd{n'P_{J}}$ with $n=6,7$ and $n'=6-12$.  Our relativistic CC calculations are complete through the fifth order of many-body perturbation theory~\cite{PorDer06Na,DerPorBel08} and include a large class of diagrams summed to all orders. As such, these are the most complete {\em ab initio} relativistic many-body calculations of $E1$ matrix elements in Cs to date. To this end we extend our earlier coupled-cluster CCSDvT calculations~\cite{Porsev2010} to the next level of computational complexity. The CCSDvT method includes  single and double excitations of electrons from the Cs Xe-like core and single, double, and triple excitations of the valence electron. Here we amend the CCSDvT method with perturbative treatment of core triples: we will use the CCSDpTvT designation for this method, with pT emphasizing the perturbative treatment of core triple excitations. 

Our compilation of $E1$ matrix elements is anticipated to be useful in a variety of applications ranging from determining atomic polarizabilities, light shifts, and magic wavelengths for laser cooling, trapping, and atom manipulation in atomic clocks~\cite{Katori1999,Ye1999,Safronova2006,Mitroy2010,Ludlow2015,Nicholson2015,Safronova2016,Cooper2018}, to evaluating the long-range interaction coefﬁcients $C_6$ and $C_8$ needed in ultra-cold collision physics~\cite{Derevianko2002,Porsev2014}, and finally to suppressing decoherence in quantum simulation, quantum information processing, and quantum sensing~\cite{Zhang2011,Goldschmidt2015}. In addition, our matrix elements can lead to more accurate theoretical determination of the $6S_{1/2}\rightarrow7S_{1/2}$ parity-violating amplitude $E_{PV}$ and transition polarizabilities. The vector transition polarizability is needed to extract $E_{PV}$ from experimental results~\cite{Wood1997}, whereas the theoretical value for $E_{PV}$ facilitates the inference of more fundamental quantities such as the electroweak Weinberg angle~\cite{RoPP2020,ERLER2013119,tan2022implications} and, thereby, improves 
precision probes of the low-energy electroweak sector of the standard model of elementary particles.

Experimentally, the $\brad{6S_{1/2}}D\ketd{6P_{J}}$ matrix elements are the most accurately known, through a variety of techniques, including time-resolved ﬂuorescence~\cite{Young1994,Patterson2015}, absorption~\cite{Rafac1999}, ground-state polarizability~\cite{Amini2003,Gregoire2015}, and photo-association spectroscopy~\cite{Derevianko2002,Bouloufa2007,Zhang2013}. The relative uncertainties in these experiments are $\sim 0.1$\%. Direct absorption measurements of $\brad{6S_{1/2}}D\ketd{7P_{J}}$ yielded results differing at the $\lesssim 1$\% level~\cite{Vasilyev2002,Antypas2013,borvak2015direct}, while a more recent experiment comparing the absorption coefficient of the $6S_{1/2}\rightarrow7P_{J}$ transitions to that of the more precisely known $6S_{1/2}\rightarrow6P_{1/2}$ line obtained $\brad{6S_{1/2}}D\ketd{7P_{1/2}}$ and $\brad{6S_{1/2}}D\ketd{7P_{3/2}}$ with uncertainties of $0.1\%$ and $0.16\%$ respectively~\cite{Damitz2019}. The $\brad{7S_{1/2}}D\ketd{7P_{J}}$ matrix elements were determined~\cite{Safronova1999} by combining a measurement of the DC Stark shift of the $7S_{1/2}$ state \cite{PhysRevA.59.R16} with the theoretical value for the ratio of matrix elements $\brad{7S_{1/2}}D\ketd{7P_{3/2}}/\brad{7S_{1/2}}D\ketd{7P_{1/2}}$ with uncertainties of $0.15\%$. An updated measurement of the $7S_{1/2}$ DC Stark shift~\cite{Toh2019Polarizabilities} reduced the uncertainties in $\brad{7S_{1/2}}D\ketd{7P_{J}}$ to $0.05\%$. The $\brad{7S_{1/2}}D\ketd{6P_{J}}$ matrix elements were determined from the measured lifetime of $7S_{1/2}$~\cite{Bouchiat1984} and the theoretical value for the ratio $\brad{7S_{1/2}}D\ketd{6P_{3/2}}/\brad{7S_{1/2}}D\ketd{6P_{1/2}}$~\cite{dzuba1989summation,BluJohSap89,Safronova1999}, with uncertainties $\sim 0.5\%$. More recently, the $7S_{1/2}$ lifetime and the ratio $\brad{7S_{1/2}}D\ketd{6P_{3/2}}/\brad{7S_{1/2}}D\ketd{6P_{1/2}}$ were remeasured with a better accuracy resulting in an improved $0.1\%$ uncertainty in the $\brad{7S_{1/2}}D\ketd{6P_{J}}$ matrix elements~\cite{Toh2019Mels}. To the best of our knowledge, high-precision experimental results for $E1$ matrix elements involving states with higher principle quantum numbers are lacking.

There were many theoretical determinations of $E1$ matrix elements in Cs (see comparisons in the later sections of this paper). Here we review the ones that are closely related to our CC methodology. In the context of atomic parity violation~\cite{Porsev2010}, the $\brad{nS_{1/2}}D\ketd{n'P_{1/2}}$ matrix elements with $n=6,7$ and $n'=6,7,8,9$ were calculated using a coupled-cluster  approach which included nonlinear singles, doubles, and valence triples (CCSDvT) with an accuracy at the level of 0.2\%. A broader study~\cite{Safronova2016} of several Cs atomic properties including lifetimes, matrix elements, polarizabilities, and magic wavelengths, presented a comprehensive list of ${nS_{1/2}}\rightarrow n'P_{1/2,3/2}$ matrix elements with $n=6-14$ and $n'=6-12$. Although $E1$ matrix elements between states with lower principle quantum numbers had estimated uncertainties around a few percent, the uncertainties in those involving higher principle quantum numbers were $\sim 20$\%.  It is worth pointing out that the CC method used in Ref.~\cite{Safronova2016} is the linearized version of the CC method limited to singles, doubles, and partial triples (SDpT), to be contrasted with our more complete CCSDpTvT method employed in the present work. The CCSDvT method of Ref.~\cite{Porsev2010} was also less complete as it did not include treatment of core triple excitations. 

This paper is organized as follows. In Sec.~\ref{Sec:Theory}, we present a summary of the methods employed in our computations of $E1$ matrix elements in Cs. Numerical results are tabulated an discussed in Sec.~\ref{Sec:Numericals}. The paper presents several  appendices where we document our methods of solving the many-body problem, such as the construction of Dirac-Hartree-Fock basis sets (Appendix~\ref{App:Finite-Basis-Set}) and Brueckner orbitals (Appendix~\ref{App:BO}), 
as well as the basis-set implementation of the random-phase approximation (Appendix~\ref{App:RPA}). These techniques were used in several earlier papers by our group and documenting them not only facilitates a reproduction of our results, but can be also useful for the community. Unless specified otherwise, atomic units, $|e|=m_e=\hbar=1$, are used.


\section{Theory}\label{Sec:Theory}

In this section, we discuss several {\em ab initio}  relativistic many-body methods we use to compute the $E1$ matrix elements. These include the lowest order  Dirac-Hartree-Fock (DHF) method, the random-phase approximation (RPA), the Brueckner-orbital (BO) method, the combined RPA(BO) method, and several levels of approximation within the CC method. We also present details of computing ``minor'' corrections: Breit, QED, semiempirical, and numerical basis-extrapolation corrections.  

\subsection{Basics}
We begin by considering the Dirac Hamiltonian of the atomic electrons propagating in the nuclear potential $\sum_i V_{\rm nuc}(r_i)$. Here, $i$ ranges over all the $N=55$ electrons of the Cs atom. The full electronic Hamiltonian $H$ may be decomposed into
\begin{equation}\label{Hamiltonian'}
    \begin{aligned}
        H &= \sum_i h_0(i) + V_c\,,\\
        h_0(i) &= c\boldsymbol{\alpha}_i\cdot{\bf p}_i+m_ec^2\beta_i\\
        &+V_{\rm nuc}(r_i)+h_W(r_i)+U(r_i)\,,\\
        V_c &=\frac12\sum_{i\neq j}\frac{e^2}{|{\bf r}_i-{\bf r}_j|} - \sum_i U(r_i)\,,
    \end{aligned}
\end{equation}
where $U(r_i)$ is chosen to be the conventional frozen-core $V^{N-1}$ DHF potential as it dramatically reduces the number of many-body perturbation theory (MBPT) diagrams. For brevity, we suppressed the positive-energy projection operators for the two-electron interactions (no-pair approximation). See the textbook~\cite{WRJBook} for further details. 

As usual, we assume that the energies $\varepsilon_i$ and orbitals $\psi_i$ of the single-electron DHF Hamiltonian $h_0$ are known. In Appendix~\ref{App:Finite-Basis-Set}, we discuss the construction of the DHF $B$-spline basis sets used in our numerical work. These basis sets approximate the spectrum of $h_0$ and are numerically complete.
With the complete spectrum of $h_0$ determined, 
the many-body eigenstates $\Psi$ of $H$ are then expanded over antisymmetrized products of the one-particle orbitals $\psi_i$. In MBPT, one obtains these eigenstates by treating the residual $e^-e^-$ interaction $V_c$ as a perturbation. Second quantization and diagrammatic techniques considerably streamline the MBPT derivations. To this end, we first express Eq.~\eqref{Hamiltonian'} in terms of $a^\dagger_i$ and $a_i$, the creation and annihilation operators associated with the one-particle eigenstate $\psi_i$ of $h_0$.  We  will follow the indexing convention that core  orbitals are labeled by the letters at the beginning of alphabet $a,b,c,\dots$, while valence electron orbitals are denoted by $v,w,\dots$, and the indices $i,j,k,\dots$ refer to an arbitrary orbital, core or excited (including valence states). The letters $m,n,p,\dots$ are reserved for those orbitals unoccupied in the core (these could be valence orbitals). 

In the second quantization formalism, the DHF Hamiltonian $H_0$ and the residual $e^-e^-$ interaction read
\begin{equation}\label{2quant}
    \begin{aligned}
        H_0 &= \sum_i\varepsilon_i N[a^\dagger_ia_i] \, ,\\
        V_c &= \frac12\sum_{ijkl}g_{ijkl}N[a^\dagger_ia^\dagger_ja_ka_l]\,,
    \end{aligned}
\end{equation}
 where $N[\cdots]$ denotes normal ordering of operator products and the Coulomb matrix elements are
\begin{equation}\label{eq:def_g}
	g_{ijkl}\equiv \int\frac{d^3{r}_1 d^3{r}_2}{\left|{\bf r}_1-{\bf r}_2\right|}\psi^\dagger_i({\bf r}_1)\psi^\dagger_j({\bf r}_2)\psi_k({\bf r}_1)\psi_l({\bf r}_2)\,.
\end{equation}

The zero-order wave function may be expressed as  $\ket{\Psi^{(0)}_v}=a^\dagger_v\ket{0_c}$, where $\ket{0_c}$ represents the filled Fermi sea of the atomic core (quasivacuum state). We are interested in a matrix element of a one-electron operator $Z=\sum_{ij}z_{ij}a^\dagger_ia_j$ between two valence many-body states $\ket{\Psi_w}$ and $\ket{\Psi_v}$, $Z_{wv}$.
The first-order contribution to the $Z_{wv}$ is given by
\begin{equation}
    Z^{(1)}_{wv}=\bra{\Psi^{(0)}_w}Z\ket{\Psi^{(0)}_v}=z_{wv}+\delta_{wv}\sum_az_{aa}\,.
\end{equation}
For the $E1$ matrix elements, the sum over core orbitals vanishes due to selection rules, and $Z^{(1)}_{wv}$ reduces to the DHF value of the matrix element $z_{wv}$.

The second-order MBPT correction to matrix elements reads
\begin{align}\label{dtwv}
        Z^{(2)}_{wv} &= \sum_{an}\frac{z_{an}\tilde{g}_{wnva}}{\varepsilon_a-\varepsilon_n-\omega}+\sum_{an}\frac{\tilde{g}_{wavn}z_{na}}{\varepsilon_a-\varepsilon_n+\omega}\,, 
    \end{align}
where $\omega\equiv\varepsilon_w-\varepsilon_v$ and $\tilde{g}_{ijkl}\equiv g_{ijkl}-g_{ijlk}$.

One may separate the third-order correction $Z^{(3)}_{wv}$ into different classes of diagrams~\cite{Johnson1996} 
\begin{equation}\label{eq:3rd corrections}
    Z^{(3)}_{wv} = Z^{\rm 3, RPA}_{wv} + Z^{\rm 3, BO}_{wv} + Z^{\rm SR}_{wv} + Z^{\rm Norm}_{wv}\,.
\end{equation}
The RPA and BO terms are discussed in Secs.~\ref{Sec:RPA} and~\ref{Sec:BO}. These corrections typically dominate the third-order contributions. Expressions for the structural radiation (SR) and normalization (Norm) terms can be found in Ref.~\cite{Johnson1996}. We do not, however, include the SR and Norm diagrams in our MBPT calculations. Their contributions, as well as higher-order ones, are more systematically accounted for in the CC approach described in Sec.~\ref{Sec:CC}.

Fourth-order diagrams $Z^{(4)}_{wv}$ have been computed in Refs.~\cite{CanDer04,DerEmmons2002}. These are subsumed in the CCSDvT method and we do not compute them explicitly. We are not aware of any work tabulating the fifth-order MBPT contributions. Due to the exploding number of  diagrams with increasing MBPT order, such contributions are more elegantly accounted for using all-order diagrammatic summation techniques. 

Diagrammatic techniques enable summing certain classes of diagrams to all orders. For example, the RPA method, discussed in Sec.~\ref{Sec:RPA}, incorporates the second-order $Z^{(2)}_{wv}$, third-order
$Z^{\rm 3, RPA}_{wv}$, and all higher-order diagrams of the similar topological structure. Similar considerations apply to the BO method (see Sec.~\ref{Sec:BO}). The CCSDvT (and, by extension, the more sophisticated CCSDpTvT) method sum even larger classes of MBPT diagrams to all orders. The CCSDvT method is complete through the fifth order of conventional MBPT~\cite{PorDer06Na,DerPorBel08}; it starts missing certain diagrams in the sixth order.

Finally, as a matter of practical implementation, the MBPT expressions, like Eq.~\eqref{dtwv}, involve summations over the core  and the excited orbitals. Each orbital $\psi_i$ is characterized by a principal quantum number $n_i$, orbital angular momentum $\ell_i$, a total angular momentum $j_i$, and its projection $m_i$. The sums over the magnetic quantum numbers $m_i$ are carried out analytically using the rules of Racah algebra. Although the sums over $j_i$ are infinite, they are restricted by angular momentum selection rules which  reduce the number of surviving terms. Moreover, the sums over total angular momenta converge well and in practice, it suffices to sum over a few lowest values of $j_i$. The sums over the principal quantum numbers $n_i$ involve, on the other hand, summing over the infinite discrete spectrum and integrating over the continuum. In the finite-basis-set method, employed in our work, these infinite summations are replaced by summations over a finite-size 
pseudospectrum~\cite{Johnson1988,Bachau2001,Shabaev2004,BELOY2008}.

The basis orbitals in the pseudospectrum are obtained by placing the atom in a sufficiently large cavity and imposing boundary conditions at the cavity wall and at the origin (see Ref.~\cite{BELOY2008} for further details on dual-kinetic-basis $B$-spline sets used in our paper). For each value of $j_i$, one then finds a discrete set of $2M$ orbitals, $M$ from the Dirac sea and the remaining $M$ with energies above the Dirac sea threshold (conventionally referred to as ``negative'' and ``positive'' energy parts of the spectrum in analogy with free-fermion solutions). This enables a straightforward implementation of the positive-energy spectrum projection operators in the no-pair approximation.

If the size of the cavity is large enough, typically about $40a_0/Z_\mr{eff}$ where $a_0$ is the Bohr radius and $Z_\mr{eff}$ is the effective charge of the core felt by the valence electrons, the low-lying basis-set orbitals map with a good accuracy to the discrete orbitals of the exact DHF spectrum obtained with the conventional finite-differencing techniques. Higher-energy orbitals do not closely match their physical counterparts due to confinement and discretization (see Sec.~\ref{Sec:Numericals} and Appendix~\ref{App:Finite-Basis-Set}). Nevertheless, since the pseudospectrum is numerically complete, in the sense that any function satisfying the boundary conditions imposed by the cavity can be expanded in terms of the basis functions, it can be used instead of the real spectrum to evaluate correlation corrections to states confined to the cavity. Theoretically, in the limit where the cavity size and the number of basis functions, $M$, go to infinity, one recovers the physical problem. The increasing computational cost associated with increasing $M$ means, however, that in practice, finite but reasonably large values of cavity radius and basis0set size are chosen and numerical errors due to these finite values are estimated by extrapolating to the infinite basis (see Sec.~\ref{sec:Basis extrapolation correction} for more details). From now on, all single-particle DHF orbitals $\psi_i$ are understood to be members of a finite basis set. Details on our construction of the $B$-spline basis set are presented in Appendix~\ref{App:Finite-Basis-Set}.

\subsection{Brueckner-orbital method}\label{Sec:BO}
Qualitatively, the BO correction accounts for a process where the valence electron charge polarizes the atomic core, inducing a dipole and higher-rank multipolar moments in the core. The valence electron is then attracted by the induced redistribution of charges in the core, reducing the size of the valence electron's orbit. This process is included in a generic model-potential formulation by  adding a relevant self-energy operator $\Sigma(\mb{r})$ to the valence electron Hamiltonian
\begin{equation}
\Sigma^\mathrm{m.p.}(\mb{r}) = -{\alpha_{c}}/(2r^{4}) \,,\label{Eq:SE-modpot}
\end{equation}
with $\alpha_{c}$ being the electric-dipole polarizability of the core. 

Note that since $\Sigma^\mathrm{m.p.}(\mb{r})$ diverges at small distances, higher multipole contributions are needed for states with low orbital angular momenta and may be more systematically accounted for in a more involved many-body formulation of the self-energy operator. For example, to second order, the matrix element of $\Sigma$ between arbitrary orbitals $i$ and $j$ is given by~\cite{JohSapBlu89} 
\begin{align}
 \Sigma^{(2)}_{ij}(\varepsilon_0)= \sum_{amn}\frac{g_{aimn}\tilde{g}_{mnaj}}{\varepsilon
_{a0}- \varepsilon_{mn}}+\sum_{abm}\frac{\tilde{g}_{miab}g_{abmj}}%
{\varepsilon_{m0}-\varepsilon_{ab}} \,, \label{Eq:SE-II}
\end{align}
where $g$ and $\tilde{g}$ are the Coulomb matrix elements as defined in Eq.\eqref{eq:def_g} and after Eq.~\eqref{dtwv}. Here we use the shorthand notation $\varepsilon_{i_1, i_2} \equiv \varepsilon_{i_1} + \varepsilon_{i_2}$, with  
$\varepsilon_0$ being some reference energy (see Appendix~\ref{App:BO} for details). We employ Eq.~\eqref{Eq:SE-II} in our calculations.
In particular, the diagonal matrix elements $\Sigma_{vv}(\varepsilon_v)$ are simply the second-order MBPT  correction to the energy  of valence state $v$. 
The multipolar expansion of $\Sigma^{(2)}(\varepsilon_v)$ in the limit of the valence electron being far away from the core recovers the model potential expression~\eqref{Eq:SE-modpot}.

The Brueckner orbitals $u$ and corresponding energies are determined by solving the eigenvalue equation with both the DHF Hamiltonian $h_0$ and the self-energy operator $\Sigma$ included:
\begin{equation}
\left(  h_0 +\Sigma^{(2)}(\varepsilon_0) \right) u=\varepsilon^\mr{BO} u \,. 
\end{equation}
Our numerical approach to solving this eigenvalue equation is discussed in Appendix~\ref{App:BO}; we solve the matrix eigenvalue problem using the DHF finite basis set. With the BO orbitals determined, the matrix element is simply $\langle u_w |z| u_v \rangle$, which includes the DHF value, third-order $Z^{\rm (3), BO}_{wv}$ contribution, and higher-order corrections.

\subsection{The random-phase approximation}\label{Sec:RPA}
Detailed introductions to the formalism of th RPA can be found in Refs.~\cite{Joh88,Amusia2013atomic}. The RPA is a linear-response theory realized in the self-consistent mean-field (DHF in our case) framework. Qualitatively, it accounts for the screening of the externally applied electric field (e.g., a driving laser field oscillating at the transition frequency $\omega$) by the core electrons. The RPA formalism is an all-order method and offers a distinct advantage of being gauge independent in computations of transition amplitudes.

The third-order RPA term in Eq.~\eqref{eq:3rd corrections} is structurally similar to $Z^{(2)}_{wv}$ and can be grouped with it. It may be shown that topologically similar diagrams  exist in higher-order MBPT corrections~\cite{fetter2003}. When all these diagrams are included, one obtains the RPA  corrections similar in form to second-order Eq.~\eqref{dtwv}. In the RPA, one first computes the ``core-to-excited'' matrix elements  $z^{\rm RPA}_{an}$ and $z^{\rm RPA}_{na}$ (RPA vertices)~\cite{Johnson1996}
\begin{equation}\label{RPA_core}
	\begin{aligned}
		z^{\rm RPA}_{an} &= z_{an} + \sum_{bm}\frac{z^{\rm RPA}_{bm}\tilde{g}_{amnb}}{\varepsilon_b-\varepsilon_m-\omega}+ \sum_{bm}\frac{\tilde{g}_{abnm}z^{\rm RPA}_{mb}}{\varepsilon_b-\varepsilon_m+\omega}\,,\\
		z^{\rm RPA}_{na} &= z_{na} + \sum_{bm}\frac{z^{\rm RPA}_{bm}\tilde{g}_{nmab}}{\varepsilon_b-\varepsilon_m-\omega}+ \sum_{bm}\frac{\tilde{g}_{nbam}z^{\rm RPA}_{mb}}{\varepsilon_b-\varepsilon_m+\omega}\,.
	\end{aligned}
\end{equation}
Once the RPA vertices
are obtained, the RPA matrix element between two valence states is given by 
\begin{equation}\label{RPA_valence-2}
	\begin{aligned}
		Z^\mr{RPA}_{wv} =\sum_{an}\frac{z^{\rm RPA}_{an}\tilde{g}_{wnva}}{\varepsilon_a-\varepsilon_n-\omega}+ \sum_{an}\frac{\tilde{g}_{wvna}z^{\rm RPA}_{na}}{\varepsilon_a-\varepsilon_n+\omega}\,.
	\end{aligned}
\end{equation}
Our numerical finite-basis-set implementation of the RPA is described in Appendix~\ref{App:RPA}.

An iterative solution of Eqs.~\eqref{RPA_core} recovers the conventional MBPT diagrams order-by-order, but starts missing contributions in the third order.
Among the missing third-order diagrams, the dominant correlation correction is usually $Z^{\rm (3), BO}_{wv}$, coming from Brueckner orbitals (see Sec.~\ref{Sec:BO}). To include the important BO correction in the RPA framework, we will also use a basis of Brueckner orbitals (instead of the DHF orbitals) in solving the RPA equations; we will denote such results as RPA(BO). The conventional RPA results using the DHF basis will be denoted RPA(DHF).   



\subsection{The coupled-cluster method}\label{Sec:CC}
 The task of accounting for higher-order MBPT corrections can be systematically carried out by means of the CC method~\cite{CoeKum60,Ciz66}, which we discuss in this section. Ultimately, we will employ the CCSDvT and the CCSDpTvT methods which are known to be complete through the fourth order of MBPT for energies and through the fifth order for matrix elements \cite{PorDer06Na,DerPorBel08}.

We begin by going back to the second-quantized form of the full electronic Hamiltonian $H$, Eq.~\eqref{2quant},
\begin{align}\label{Eq:SecQuantH}
        H &=  H_0 + G \nonumber\\
          &= \sum_i \varepsilon_i N[a^\dagger_i a_i] + \frac12\sum_{ijkl} g_{ijkl} N[a^\dagger_i a^\dagger_j a_l a_k ] \,.
\end{align}

It may be shown that the exact many-body eigenstate $\ket{\Psi_v}$ of $H$ may be represented as 
\begin{align}\label{Eq:PsivOmega}
        |\Psi_v\rangle &=  N[ \exp(K) ]\, |\Psi^{(0)}_v\rangle\nonumber\\
                    &= \left( 1 + K + \frac{1}{2!} N[K^2] + \ldots \right)
\, |\Psi^{(0)}_v\rangle \, ,
\end{align}
where $|\Psi^{(0)}_v \rangle$ is the lowest-order DHF state and the cluster operator $K$ is expressed in terms of connected diagrams of the wave operator \cite{lindgren2012atomic}
\begin{align}\label{Eq:KCCSDvT}
         K &= S_c + D_c + T_c + S_v + D_v + T_v + \ldots \nonumber \\
           &= \sum_{ma}\rho_{ma}a^\dagger_m a_a + \frac{1}{2!}\sum_{mnab}\rho_{mnab}a^\dagger_ma^\dagger_na_ba_a\nonumber\\
           &+ \frac{1}{3!}\sum_{mnrabc}\rho_{mnrabc}a^\dagger_ma^\dagger_na^\dagger_ra_ca_ba_a\nonumber\\
           &+\sum_{m\neq v}\rho_{mv}a^\dagger_m a_v+ \frac{1}{2!}\sum_{mna}\rho_{mnva}a^\dagger_ma^\dagger_na_aa_v\nonumber\\
           &+ \frac{1}{3!}\sum_{mnrab}\rho_{mnrvab}a^\dagger_ma^\dagger_n a^\dagger_ra_ba_aa_v+\ldots\,.
    \end{align}
Here $S_v$, $D_v$, and $T_v$ ($S_c$, $D_c$, and $T_c$) are the valence (core) singles, doubles, and triples, expressed in terms of the creation and annihilation operators $a^\dagger_i$ and $a_i$. By substituting Eqs.~\eqref{Eq:PsivOmega} and~\eqref{Eq:KCCSDvT} into  the Bloch equation  specialized for univalent systems \cite{DerEmmons2002}, we obtain a set of coupled algebraic equations for the cluster amplitudes $\rho$. 
We solve the CC equations numerically using finite basis sets, obtaining, as a result, the cluster amplitudes $\rho$ and the correlation corrections to the valence electron energies $\delta E_v$.

The explicit form of these equations depends on the level of approximation at which one chooses to operate. For example, one may truncate the expansion \eqref{Eq:KCCSDvT} at doubles and the expansion~\eqref{Eq:PsivOmega} at the term linear in $K$. 
The resulting linear singles-doubles approximation is conventionally labeled ``SD''. If one chooses to retain only singles and doubles but all nonlinear terms in Eq.~\eqref{Eq:PsivOmega}, one obtains the nonlinear singles-doubles approximation, labeled ``CCSD''. Contributions from core triples may be partially accounted for by considering their perturbative effects on core singles and doubles, corresponding to the ``CCSDpT" method. In this work, we will employ both the ``CCSDvT'' and ``CCSDpTvT'' methods, which include the valence triples, corresponding to the term $T_v$ in Eq.~\eqref{Eq:KCCSDvT}, on top of the core CCSD and CCSDpT. The topological structure and explicit form of the Bloch equations in these approximations may be found in Refs.~\cite{PalSafJoh07,DerPorBel08}.

Once the cluster amplitudes $\rho$ and thus the many-body wave functions for two valence states $v$ and $w$ have been obtained, one may evaluate the $E1$ matrix element between $w$ and $v$ using
\begin{equation}\label{eq:Dwv}
    D_{wv}=\frac{\bra{\Psi_w}\sum_{ij}d_{ij}a^\dagger_ia_j\ket{\Psi_v}}{\sqrt{\braket{\Psi_w|\Psi_w}\braket{\Psi_v|\Psi_v}}}\,,
\end{equation}
where $d_{ij}\equiv\bra{i}d\ket{j}$ are the single-electron $E1$ matrix elements. The corresponding expressions for different contributions to $D_{wv}$ are given in Refs.~\cite{BluJohSap89} and~\cite{PorDer06Na}. Note that these expressions include only linear single, linear double, and linear triple contributions to $D_{wv}$. Additional modifications to $D_{wv}$ due to the nonlinear single and double terms in the CC wave functions are accounted for by the ``dressing'' of lines and vertices~\cite{DerPor05}. See Sec.~\ref{sec:Dressing} for more details.

\subsection{Other corrections}
\subsubsection{Semiempirical scaling}\label{sec:Semiempirical scaling}
Since our most complete CCSDpTvT method is still an approximation, we miss certain correlation effects (due to our perturbative treatment of core triples and omission of core and valence quadruple and higher-rank excitations). This is the cause of the
difference between the computed and experimental energies.
To partially account for the missing contributions in calculations of matrix elements, we additionally correct the CCSDpTvT wave functions using a semiempirical procedure suggested in Ref.~\cite{Blundell1992}. 

This approach is based on the observation that there exists a nearly linear correlation between the variations of correlation energies and matrix elements in different approximations. This linear dependence is due to the effect of self-energy correction, which gives rise to one of the dominant chains of diagrams present in both matrix elements and energies. For example, for triple excitations, the corrections $S_v[T_v]$ and $\delta E_v[T_v]$ (in the notations of Ref.~\cite{PorDer06Na}) arise from the same diagram and the modiﬁcation of singles due to triples ($S_v[T_v]$) propagates into the calculation of the matrix element. 

More specifically, a dominant contribution to the majority of matrix elements comes from the BO-like term involving valence singles (following the notation of Ref.~\cite{BluJohSap89})
\begin{equation}
    Z_{wv}^{(c)}=\sum_mz_{wm}\rho_{mv}+z_{mv}\rho_{mw}\,.
\end{equation}
One may connect the CC $Z_{wv}^{(c)}$ diagram to a BO-basis matrix element 
$\langle u_w | z | u_v \rangle$ via $u_v = \sum_m \rho_{mv} \psi_m$, with $\rho_{mv}$ being the expansion coefficients over DHF basis set $\{\psi_m\}$ (see Sec.~\ref{Sec:BO}). 
Missing corrections to $Z_{wv}^{(c)}$ due to higher-rank CC excitations may be partially accounted for by improving the values of the valence single coefﬁcients $\rho_{mv}$. This is achieved by noting that the correlation energy and single amplitudes are closely related. Indeed, the self-energy operator $\Sigma$ defined in Sec.~\ref{Sec:BO} is connected to the valence singles via
\begin{equation}\label{eq:Singles}
   (\varepsilon_v-\varepsilon_m
    +\delta E_v^{\rm CC})\rho_{mv} =  \Sigma_{mv}\,,
\end{equation}
where $\delta E_v^{\rm CC}$ is the correlation energy computed at the given level of CC approximation (and approaches true value of correlation energy in the complete, yet practically unattainable for Cs, treatment). Notice that the role of $\delta E_v^{\rm CC}$ on the left-hand side of Eq.~\eqref{eq:Singles} is suppressed as typically $|\delta E_v^{\rm CC}| \ll |\varepsilon_v-\varepsilon_m|$.
More importantly, the diagonal matrix element of $\Sigma$ is the correlation correction to the energy of valence state $v$, $\delta E^{\rm CC}_v=\Sigma_{vv}$. As a result, contributions from higher-order diagrams to the right-hand side of Eq.~\eqref{eq:Singles} are similar to those to the correlation energy. 

This observation suggests rescaling the valence single coefﬁcients as~\cite{BluJohSap89}
\begin{equation}\label{eq:scaling}
    \rho'_{mv}=\rho_{mv}\frac{\delta E_v^{\rm expt}}{\delta E_v^{\rm CC}}\,,
\end{equation}
where $\delta E_v^{\rm exp}$ and $\delta E_v^{\rm CC}$ are the experimental and computed correlation energies at a given level of CC approximation, respectively. Note that a consistent deﬁnition of the experimental correlation energies requires removing the Breit, QED, and basis extrapolation corrections (see Sec.~\ref{sec:Basis extrapolation correction} below) from the experimental energy, i.e.,
\begin{align}
    \delta E_v^{\rm expt}&=E_v^{\rm expt}-E_v^{\rm DHF}-\delta E_v^{\rm Breit}\nonumber\\
    &-\delta E_v^{\rm QED}-\delta E_v^{\rm extrapol}\,.
\end{align}
We have removed basis extrapolation correction $\delta E_v^{\rm extrapol}$ from energy because the extrapolation correction to matrix elements  is computed separately.

It is worth emphasizing, however, that the linear scaling of matrix elements with correlation energy is only approximate and can be used in the semiempirical ﬁts only to a certain accuracy. For example, as will be discussed in Sec.~\ref{Sec:Numericals}, scaling at the singles and doubles (SD) level generally does not necessarily produce a result compatible with that obtained using a more complete method, say CCSDvT or CCSDpTvT, partially because these methods include additionally a direct valence triples correction to matrix elements (systematic shifts in the language of experimental physics).
Similarly, the self-energy corrections do not affect the dressing of matrix elements (see Sec.~\ref{sec:Dressing}). Nor can it capture the distinctively different QED corrections to the energies and matrix elements. We refer the reader to Ref.~\cite{DerPorBel08} for further justification and discussion of caveats of the semiempirical scaling in the CC method context.

\subsubsection{Dressing of matrix elements}\label{sec:Dressing}

Once one has obtained the CC amplitudes by solving the CC equations [and rescaling the single amplitudes as per Eq.~\eqref{eq:scaling}], one may proceed to computing the matrix element $D_{wv}$ by substituting Eq.~\eqref{Eq:PsivOmega} into Eq.~\eqref{eq:Dwv}. 
Notice that the CC wavefunction~\eqref{Eq:PsivOmega} includes an exponential of the cluster operator $K$,
$|\Psi_v\rangle = \left( 1 + K + \frac{1}{2!} N[K^2] + \ldots \right) \, a_v^\dagger |0_v\rangle$. Dressing of matrix elements refers to the inclusion of nonlinear terms in the above expansion into the computations of matrix elements. In general, there is an infinite number of such contributions even if the cluster operator $K$ is truncated 
at a certain number of excitations.
%
%
%
%
A procedure~\cite{DerPor05} for partially accounting for nonlinear contributions to matrix elements proceeds by expanding the product $C^\dagger C$ of the core cluster amplitude $C=S_c+D_c+\ldots$ into a sum of $n$-body insertions. Among these, the one- and two-body terms give the dominant contributions. The former generates diagrams with attachments to free particle and hole lines while the latter generates diagrams with two free particle (hole) lines being coupled. Summing these diagrams to all orders gives the particle and hole line dressing as well as the two-particle and two-hole RPA-like dressing. The summations over the resulting infinite series of diagrams are implemented by solving iteratively a set of equations for the expansion coefficients of the line and RPA-like dressing amplitudes. For more details, see Ref.~\cite{DerPor05}.



\subsubsection{Breit corrections}
The Breit interaction corrections to the $E1$ matrix elements and energies are computed using the MBPT formalism and numerical approaches documented in Ref.~\cite{derevianko_2001_Breit_NeutronSkin}. Briefly, we generate two basis sets, one using the conventional $V^{N-1}$ DHF potential and the other, the $V^{N-1}$ Breit-DHF potential. The Breit-DHF potential, in addition to the DHF potential, includes the one-body part of the Breit interaction between the atomic electrons in a mean-field fashion. The generation of the DHF basis sets is discussed in Appendix~\ref{App:Finite-Basis-Set}; we use identical basis-set parameters for both the DHF and Breit-DHF sets. We then carry out the RPA(BO) calculations using these two distinct basis sets (see Sec.~\ref{Sec:RPA}). For the Breit-DHF basis set, we additionally include the two-body (residual) Breit interaction on an equal footing with the residual Coulomb interaction. The Breit correction then is simply the difference between the two  RPA(BO) results. Our numerical results are consistent with Breit corrections to $E1$ matrix elements listed in Table III of Ref.~\cite{Porsev2010}.

\subsubsection{QED corrections}\label{sec:QED}
The QED corrections to $E1$ matrix elements were calculated using the radiative potential method, as developed in Refs.~\cite{Flambaum2005,Ginges2016}.
In that approach, an approximate local potential is included into the atomic Hamiltonian, which accounts for dominant vacuum polarization and electron self-energy effects. The potential is included into the DHF equation and gives an important contribution known as core relaxation, which is particularly important for states with $\ell>0$~\cite{Derevianko2004,Ginges2016,Fairhall2023}. The corrections for $\brad{6,7S}D\ketd{6,7P_J}$ were published recently in Ref.~\cite{Fairhall2023}. The authors of Ref.~\cite{Fairhall2023} have provided us with their results for the QED corrections to both energies and $E1$ matrix elements. Note that the so-called vertex corrections to $E1$ matrix elements were not included in the calculations of Ref.~\cite{Fairhall2023}. These corrections are expected to be small, due to the ``low-energy theorem''~\cite{Flambaum2005}, and account for up to a quarter of the total QED corrections in Cs. As a result, the estimated uncertainty associated with the use of the radiative potential for evaluating QED corrections to $E1$ amplitudes is at 25\%.

\subsubsection{Basis extrapolation correction}\label{sec:Basis extrapolation correction}
We perform our calculations using a basis comprising single-particle atomic orbitals with a finite number of orbital angular momenta and a finite number of $B$-spline basis-set functions for each partial wave. The basis functions are also confined within a cavity of finite, albeit large, radius. Although the finiteness of the basis greatly facilitates the efficiency of numerical procedures, it inevitably introduces some numerical errors into the final results compared to the ideal case where the cavity size, the angular momenta of the orbitals, and the number of splines per partial wave tend to infinity. For a particular atomic property $f$ ($f$ can be the removal energy or the electric-dipole matrix element), the finite-basis corrections to $f$ may be estimated by approximating $f$ with a function of the maximum orbital angular momentum $\ell_{\rm max}$, the number of splines per partial wave $M$, and the cavity radius $R_{\rm cav}$, and then extrapolating $f\left(\ell_{\rm max},M,R_{\rm cav}\right)$ to the case where all three parameters approach infinity.

We determine the dependence on $\ell_{\rm max}$ by computing $f$ in the relatively computationally inexpensive SD approximation with varying $\ell_{\rm max}$ while keeping $M$ and $R_{\rm cav}$ fixed. We then form the quantities $g(\ell_{\rm max})\equiv f\left(\ell_{\rm max}\right)-f\left(\ell_{\rm max}-1\right)$ which represents how much $f$ varies as $\ell_{\rm max}$ increases by one unit. The function $g(l)$ is estimated by fitting to $g(l)=l^{-4}(a+b/l+c/l^2)$ with fitting parameters $a$, $b$, and $c$. The correction $\delta f_{\ell_{\rm max}}=f(\infty)-f(\ell_{\rm max})$ is then approximated by $\sum_{l=\ell_{\rm max}+1}^\infty g(l)$. Similarly, the dependence on $M$ (or $R_{\rm cav}$) is determined by computing $f$ in the SD approximation with varying $M$ (or $R_{\rm cav}$) and while keeping $\ell_{\rm max}$ and $R_{\rm cav}$ (or $M$) fixed. The difference $h(M)\equiv f(M)-f(M-\Delta M)$ [or $j(R_{\rm cav})\equiv f(R_{\rm cav})-f(R_{\rm cav}-\Delta R_{\rm cav})$] is formed and fitted to $h(M)=M^{-4}(a+b/M+c/M^2)$ [or $j(R)=R^{-4}(a+b/R+c/R^2)$]. The corrections $\delta f_{M}=f(\infty)-f(M)$ and $\delta f_{R_{\rm cav}}=f(\infty)-f(R_{\rm cav})$ are approximated by $h(M+\Delta M)+h(M+2\Delta M)+\ldots$ and $j(R_{\rm cav}+\Delta R_{\rm cav})+j(R_{\rm cav}+2\Delta R_{\rm cav})+\ldots$, respectively. The total basis extrapolation correction is the sum of the three individual corrections, i.e.,
\begin{equation}
    \delta f_{\rm basis}=\delta f_{\ell_{\rm max}}+\delta f_{M}+\delta f_{R_{\rm cav}}\,.
\end{equation}
We point out that $\delta f_{\ell_{\rm max}}$ is often at least an order of magnitude larger than $\delta f_{M}$ and $\delta f_{R_{\rm cav}}$ for our basis sets.

\section{Numerical results and Discussions}\label{Sec:Numericals}

In the previous section, we have presented the theoretical basis for several methods employed in our estimates of the $E1$ matrix elements $\brad{nS_{1/2}}D\ketd{n'P_{J}}$ with $n=6,7$ and $n'=6-12$ in Cs. Numerical results for energies are presented in Tables~\ref{tab:Energy_S1}-\ref{tab:Energy_P3}, those for $E1$ matrix elements in Tables~\ref{tab:E1diople_6S1nP1}-\ref{tab:E1diople_7S1nP3} and those for the normalized ratios $\xi_{n,n'}$ in Tables~\ref{tab:ratio_6} and~\ref{tab:ratio_7}. In these tables, the final results of our computations are taken as the CCSDpTvT values with all the additional corrections (scaling, dressing, Breit, QED, and basis extrapolation) added.

In our calculations, we employed a dual-kinetic-balance $B$-spline basis set which numerically approximates a complete set of single-particle atomic orbitals. In order to accurately approximate orbitals with high principle quantum numbers, we use a large basis set containing $M=60$ basis functions for each partial wave. The basis functions are generated in a cavity of radius $R_{\rm cav}=250$ a.u. which ensures that high-$n$ orbitals, whose maxima lie far away from the origin, are not disturbed by the cavity. We test the suitability of our one-electron basis functions by comparing their corresponding energies, hyperfine structure constants, and $E1$ matrix elements with those obtained using the finite-difference solutions of the free, i.e., without cavity, DHF equations (see Appendix \ref{App:Finite-Basis-Set}). All differences are $\lesssim 0.01\%$. We note that the basis set used in Ref.~\cite{Safronova2016} yielded single-electron $E1$ matrix elements for high-$n'$ states differing from the DHF values at the level of $1\%$ . Since Ref.~\cite{Safronova2016} estimated the final uncertainties in these matrix elements at the level of $20\%$, the unphysical nature of the basis employed is irrelevant. For the purpose of our work, however, ensuring that the high-$n'$ basis functions faithfully represent their physical counterparts is essential for controlling numerical accuracy.

Basis functions with $\ell_{\rm max}\leq 7$ partial waves are used for the RPA, while in the BO and RPA(BO) approches, only partial waves with $\ell_{\rm max}\leq5$ are included due to the higher computational costs. In the CC approaches, basis functions with $\ell_{\rm max}\leq 5$ are used for single and double excitations, while for triples, we employ a more limited set of functions with $\ell_{\rm max}\leq 4$. Additionally, excitations from core subshells $[4s,\ldots,5p]$ are included in the calculations for triples, while excitations from core subshells $[1s,\ldots,3s]$ are neglected. For each partial wave, only 52 out of 60 splines are included. Basis set extrapolation corrections to inﬁnitely large $\ell_{\rm max}$, $M$, and $R_{\rm cav}$ are added separately. To estimate these corrections, SD calculations with $\ell_{\rm max}=4,5,6,7$, $M=40,60,80,100$, and $R_{\rm cav}=100, 150, 200, 250$ a.u. are performed as discussed in Sec.~\ref{sec:Basis extrapolation correction}.

We carried out computations  on a nonuniform grid defined as $\ln(r[i]/r_0+1)+a_g r[i]=(i-1){h}$ with 500 points. With $r_0=6.96\times10^{-6}$ a.u., $a_g=0.50528$, and $h=2.8801\times10^{-1}$ a.u., there are 11 points inside the ${}^{133}$Cs nucleus. The nuclear charge distribution is approximated by a Fermi distribution $\rho_{\rm nuc}(r)=\rho_0/(1+\exp[(r-c)/a])$, where $\rho_0$ is a normalization constant. For ${}^{133}$Cs, we used $c=5.6748 \, {\rm fm}$ and $a =0.52338 \, {\rm fm}$.



Our results for the removal energies are presented in Tables~\ref{tab:Energy_S1},~\ref{tab:Energy_P1}, and~\ref{tab:Energy_P3}. It may be observed that our calculations consistently underestimate the removal energies. The theory-experiment agreement improves with increasing principal quantum numbers, as expected, since orbitals with higher $n$ do not penetrate the atomic core as strongly as those with lower $n$. Such a qualitative argument becomes more explicit by considering the expectation values (first-order corrections) of the model-potential self-energy operator~\eqref{Eq:SE-modpot}, $\Sigma^\mathrm{m.p.}(\mathbf{r}) \propto 1/r^4$. The uncertainties in the final results are taken as quadrature sums of those in the CC approximation and the Breit, QED,
and basis extrapolation corrections. We estimate the systematic uncertainties in the CC approximation as the difference between the CCSDpTvT and CCSDpT values, representing higher-order terms that are missed by the CCSDpTvT approximation. The relative uncertainties in the QED corrections are estimated at the level of 25\%~\cite{Fairhall2023}. We take a conservative estimate of the uncertainties in the Breit and basis extrapolation corrections at 50\%.

\begin{table}[ht]
    \centering
    \begin{tabular}{ld{2}d{2}}
    \hline\hline
    \multicolumn{1}{c}{}& \multicolumn{1}{c}{$6S_{1/2}$}&\multicolumn{1}{c}{$7S_{1/2}$}\\
    \hline
    DHF                     & 27954 & 12112 \\ 
    BO                      & 31804 & 13023 \\
    SD                      & 31844 & 12944 \\
    CCSD                    & 31459 & 12884 \\
    CCSDpT                  & 31486 & 12889 \\
    CCSDvT                  & 31305 & 12852 \\
    CCSDpTvT                & 31332 & 12858 \\
    \hline
    \multicolumn{3}{c}{Other corrections}\\
    Breit                   & 2.6   & 0.3 \\
    QED                     &-21.5  &-5.0 \\
    Basis extrapolation     & 12.6  & 2.7 \\
    \hline
    Final result            & 31326(154) & 12856(31) \\
    Uncertainty (\%)        & 0.49 & 0.24 \\
    \hline
    Experiment~\cite{NIST}              
                            & 31406 & 12871 \\
    \hline
    Difference (\%)          &-0.26  &-0.12 \\
    Difference ($\sigma$)    &-0.52  &-0.48 \\
    \hline\hline
    \end{tabular}
    \caption{Removal energies (in ${\rm cm}^{-1}$) of $6S_{1/2}$ and $7S_{1/2}$ states in Cs in various approximations: (i) Dirac-Hartree-Fock (DHF),  (ii) Brueckner orbitals (BO), (iii) linearized coupled-cluster approximation with singles and doubles (SD), (iv) coupled-cluster approximation with singles and doubles (CCSD), (v) coupled-cluster approximation with singles and doubles and perturbative treatment of core triples (CCSDpT), (vi) coupled-cluster approximation with singles and doubles and full treatment of valence triples (CCSDvT), and (vii) the most sophisticated 
    coupled-cluster approximation with singles and doubles, perturbative treatment of core triples, and full treatment of valence triples (CCSDpTvT). The final results are obtained by adding CCSDpTvT and ``Other corrections'' entries.}
    \label{tab:Energy_S1}
\end{table}

\begin{table*}[ht]
    \centering
    \begin{tabular}{ld{2}d{7}d{6}d{6}d{6}d{6}d{6}}
    \hline\hline
    & \multicolumn{1}{c}{$6P_{1/2}$}&\multicolumn{1}{c}{$7P_{1/2}$}&\multicolumn{1}{c}{$8P_{1/2}$} &\multicolumn{1}{c}{$9P_{1/2}$}&\multicolumn{1}{c}{$10P_{1/2}$}& \multicolumn{1}{c}{$11P_{1/2}$}& \multicolumn{1}{c}{$12P_{1/2}$}\\
    \hline
    DHF                     & 18791 & 9222.6 & 5513.3 & 3671.4 & 2621.1 & 1965.3 & 1528.2 \\
    BO                      & 20290 & 9681.2 & 5718.4 & 3781.4 & 2687.1 & 2008.3 & 1558.4 \\
    SD                      & 20413 & 9686.7 & 5716.4 & 3779.1 & 2685.3 & 2006.6 & 1556.4 \\
    CCSD                    & 20230 & 9641.9 & 5698.0 & 3769.6 & 2679.7 & 2003.1 & 1554.0 \\
    CCSDpT                  & 20238 & 9644.5 & 5699.1 & 3770.3 & 2680.1 & 2003.3 & 1554.2 \\
    CCSDvT                  & 20187 & 9630.7 & 5693.2 & 3767.2 & 2678.3 & 2002.2 & 1553.4 \\
    CCSDpTvT                & 20195 & 9633.2 & 5694.4 & 3767.8 & 2678.7 & 2002.4 & 1553.6 \\
    \hline
    \multicolumn{8}{c}{Other corrections}\\
    Breit                   & -7.1  & -2.5   & -1.1   & -0.6   & -0.4   & -0.2   & -0.2   \\
    QED                     &  1.1  &  0.4   &  0.2   &  0.1   &  0.1   &  0.0   &  0.0   \\
    Basis extrapolation     &  3.5  &  1.0   &  0.5   &  0.2   &  0.1   &  0.1   &  0.1   \\
    \hline
    Final result            & 20193(43) & 9632.1(11.4) & 5694.0(4.7) & 3767.5(2.5) & 2678.5(1.4) & 2002.3(0.9) & 1553.5(0.6) \\
    Uncertainty (\%)         & 0.21 & 0.12 & 0.08 & 0.07 & 0.05 & 0.05 & 0.04 \\
    \hline
    Experiment~\cite{NIST}              
                            & 20228 & 9641.1 & 5697.6 & 3769.5 & 2679.7 & 2003.0 & 1554.0 \\
    \hline
    Difference (\%)          & -0.18 & -0.09  & -0.06  & -0.05  & -0.04  & -0.03  & -0.03  \\
    Difference ($\sigma$)          & -0.82 & -0.79  & -0.76  & -0.79  & -0.85  & -0.77  & -0.82  \\
     \hline\hline
    \end{tabular}
    \caption{Removal energies (in ${\rm cm}^{-1}$) of $nP_{1/2}$ states for $n=6-12$ in Cs in various approximations. See Table~\ref{tab:Energy_S1} caption for explanation of entries.}
    \label{tab:Energy_P1}
\end{table*}

\begin{table*}[ht]
    \centering
    \begin{tabular}{ld{2}d{7}d{6}d{6}d{6}d{6}d{6}}
    \hline\hline
    \multicolumn{1}{c}{}& \multicolumn{1}{c}{$6P_{3/2}$}&\multicolumn{1}{c}{$7P_{3/2}$}&\multicolumn{1}{c}{$8P_{3/2}$} &\multicolumn{1}{c}{$9P_{3/2}$}&\multicolumn{1}{c}{$10P_{3/2}$}& \multicolumn{1}{c}{$11P_{3/2}$}& \multicolumn{1}{c}{$12P_{3/2}$}\\
    \hline
    DHF                     & 18389 & 9079.2 & 5445.9 & 3634.4 & 2598.7 & 1950.7 & 1518.2 \\
    BO                      & 19733 & 9495.6 & 5633.4 & 3735.4 & 2659.4 & 1990.2 & 1545.7 \\
    SD                      & 19835 & 9500.5 & 5631.8 & 3733.4 & 2657.8 & 1988.8 & 1544.3 \\
    CCSD                    & 19669 & 9458.7 & 5614.3 & 3724.5 & 2652.6 & 1985.5 & 1542.0 \\
    CCSDpT                  & 19676 & 9461.0 & 5615.4 & 3725.0 & 2652.9 & 1985.7 & 1542.1 \\
    CCSDvT                  & 19632 & 9448.5 & 5610.0 & 3722.2 & 2651.2 & 1984.6 & 1541.4 \\
    CCSDpTvT                & 19639 & 9450.8 & 5611.0 & 3722.7 & 2651.6 & 1984.9 & 1541.6 \\
    \hline
    \multicolumn{8}{c}{Other corrections}\\
    Breit                   & -0.8  & -0.4   & -0.2   & -0.1   & -0.1   & 0.0    & 0.0    \\
    QED                     &  0.1  &  0.0   &  0.0   &  0.0   &  0.0   & 0.0    & 0.0    \\
    Basis extrapolation     &  3.5  &  1.1   &  0.5   &  0.3   &  0.1   & 0.1    & 0.1    \\
    \hline
    Final result            & 19642(37) & 9451.5(10.2) & 5611.3(4.4) & 3722.9(2.3) & 2651.6(1.3) & 1985.0(0.8) & 1541.7(0.5) \\
    Uncertainty (\%) & 0.19 & 0.11 & 0.08 & 0.06 & 0.05 & 0.04 & 0.03 \\
    \hline
    Experiment~\cite{NIST}             
                            & 19674 & 9460.1 & 5615.0 & 3724.8 & 2652.8 & 1985.6 & 1542.1 \\
    \hline
    Difference (\%)          & -0.16 & -0.09 & -0.07 & -0.05 & -0.05 & -0.03 & -0.03 \\
    Difference ($\sigma$)          & -0.87 & -0.84 & -0.84 & -0.82 & -0.92 & -0.75 & -0.80 \\
     \hline\hline
    \end{tabular}
    \caption{Removal energies (in ${\rm cm}^{-1}$) of $nP_{3/2}$ states for $n=6-12$ in Cs in various approximations. See Table~\ref{tab:Energy_S1} caption for explanation of entries.}\label{tab:Energy_P3}
\end{table*}

Our results for the reduced $E1$ matrix elements $\brad{nS_{1/2}}D\ketd{n'P_J}$ are compiled in Tables~\ref{tab:E1diople_6S1nP1},~\ref{tab:E1diople_6S1nP3},~\ref{tab:E1diople_7S1nP1}, and~\ref{tab:E1diople_7S1nP3}. The uncertainties in the final results are taken as quadrature sums of those in the scaling, Breit, QED, and basis extrapolation corrections. We assume that the uncertainty in the scaling correction is half its value, representing higher-order terms that are missed by the CCSDpTvT approximation. We assume that the uncertainties in matrix element dressings are already accounted for in the scaling uncertainties. Indeed, at any level of the CC approximation, the dressing corrections account for a large class of the most important diagrams arising from nonlinear CC contributions to matrix elements. As a result, it is expected that missing contributions to matrix elements come from neglecting higher-order diagrams in computing the CC amplitudes themselves, i.e., terms (partially) accounted for by the semiempirical scaling. Again, the relative uncertainties in the QED corrections are estimated at the level of 25\%~\cite{Fairhall2023} and we assume a conservative estimate of the uncertainties in the Breit and basis extrapolation corrections at 50\%. We note that since the QED, Breit, and basis extrapolation corrections are generally smaller than the semiempirical scaling ones, the uncertainties in the latter make up most of the overall uncertainty budget. The relative roles of these ``other corrections'' to the uncertainties of our results may be understood further by examining their contributions to the matrix elements themselves. 

\begin{figure}
    \centering
    \includegraphics[width=\columnwidth]{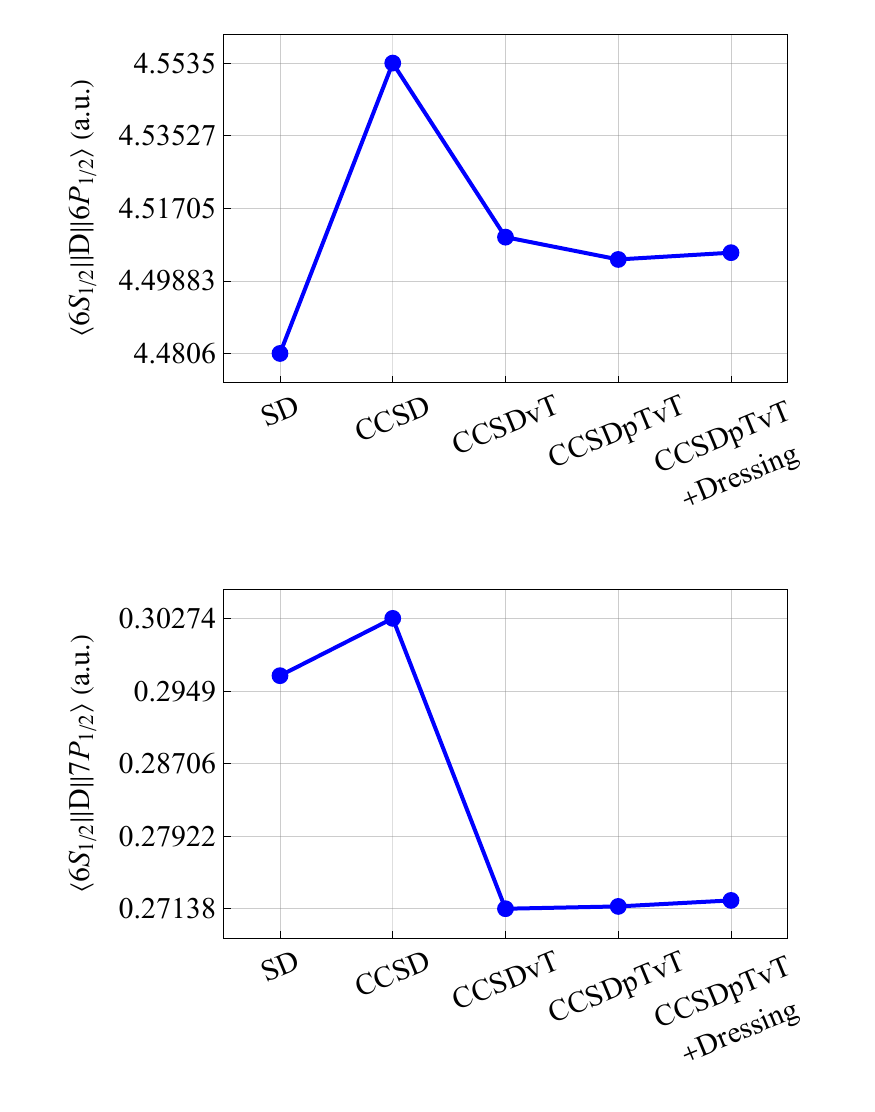}
    \caption{Convergence patterns for the $\brad{6S_{1/2}}D\ketd{nP_{1/2}}$ matrix elements with increasing complexity of the coupled-cluster method. The pattern for $n\geq8$ is similar to that for $n=7$. For all $6\leq n\leq12$, the convergence pattern for $\brad{6S_{1/2}}D\ketd{nP_{3/2}}$ is similar to that of $\brad{6S_{1/2}}D\ketd{nP_{1/2}}$.}\label{fig:convpatt6s1np1}
\end{figure}

\begin{figure}
    \centering
    \includegraphics[width=\columnwidth]{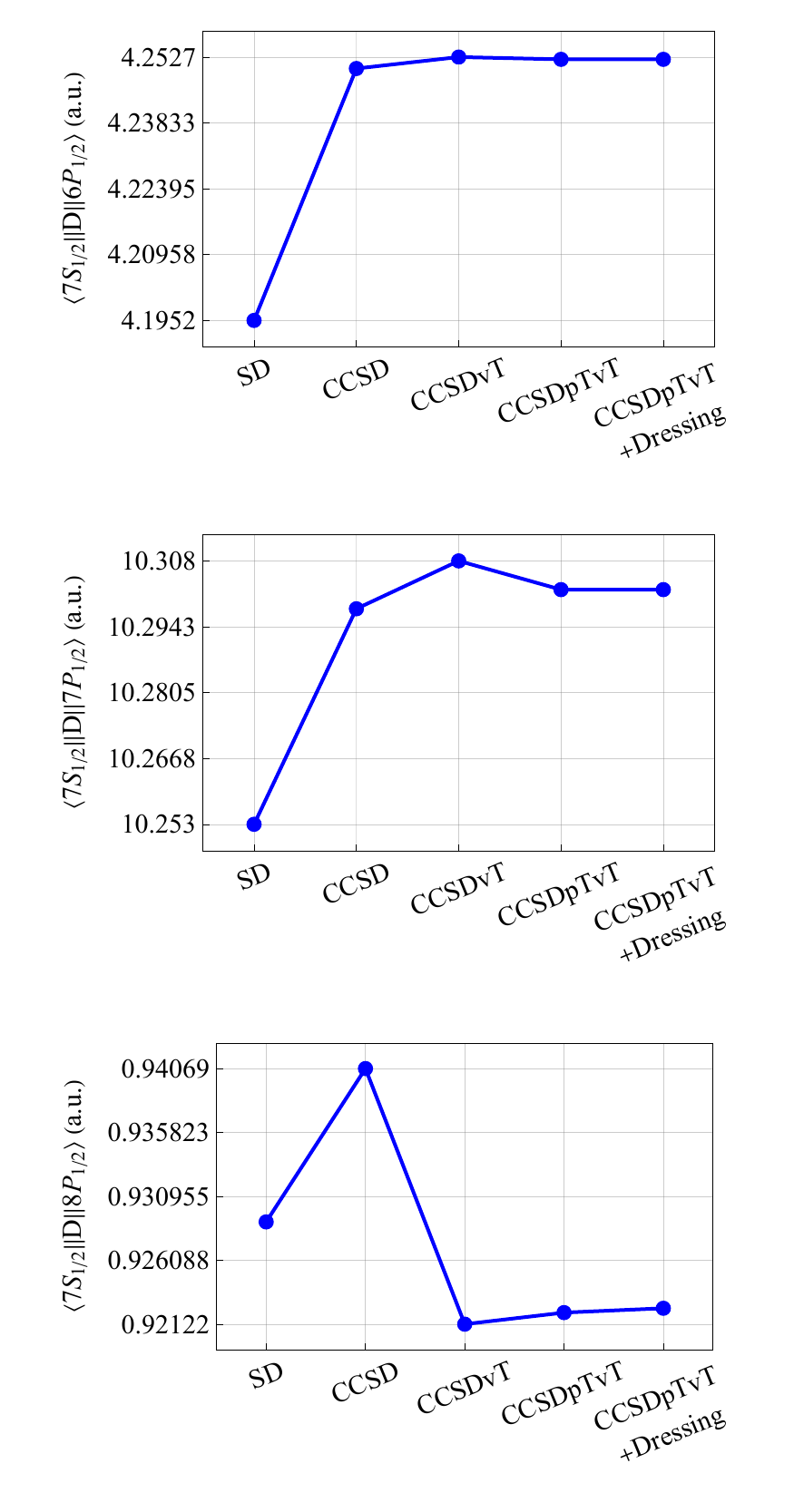}
    \caption{Convergence patterns for the $\brad{7S_{1/2}}D\ketd{nP_{1/2}}$ matrix elements with increasing complexity of the coupled-cluster method. The pattern for $n\geq9$ is similar to that for $n=8$. For all $6\leq n\leq12$, the convergence pattern for $\brad{7S_{1/2}}D\ketd{nP_{3/2}}$ is similar to that of $\brad{7S_{1/2}}D\ketd{nP_{1/2}}$.}\label{fig:convpatt7s1np1}
\end{figure}

The higher-order terms that are missed by the CCSDpTvT approximation, represented by the scaling corrections, are quite small, as may be expected if one considers the convergence patterns of the matrix elements with increasing complexity of CC approximations. Indeed, Figs.~\ref{fig:convpatt6s1np1} and \ref{fig:convpatt7s1np1} show the diminishing of contributions from higher-order diagrams: although nonlinear core singles and doubles and valence triples contribute significantly, core triples and dressing effects are generally small. More specifically, for $\brad{6S_{1/2}}D\ketd{nP_{1/2}}$, core triples account for $\lesssim1\%$ and dressings $\lesssim2\%$ of the final results. For $\brad{6S_{1/2}}D\ketd{nP_{3/2}}$, their contributions are $\lesssim0.1\%$ and $\lesssim0.5\%$, respectively. For $\brad{7S_{1/2}}D\ketd{nP_{1/2}}$, the core triples contribution is $\lesssim0.2\%$ and dressings are $\lesssim0.1\%$. For $\brad{7S_{1/2}}D\ketd{nP_{3/2}}$, both contributions are $\lesssim0.1\%$. Scaling accounts for up to $7\%$ of the final result in $\brad{6S_{1/2}}D\ketd{nP_{1/2}}$, up to $2\%$ in $\brad{6S_{1/2}}D\ketd{nP_{3/2}}$, and up to $0.6\%$ in $\brad{7S_{1/2}}D\ketd{nP_{J}}$. Note also that although not shown in Tables~\ref{tab:E1diople_6S1nP1}-\ref{tab:E1diople_7S1nP3}, we also computed the scaling corrections to the CCSDvT matrix elements and, reassuringly, found that the CCSDpTvT scaling corrections are generally smaller than the CCSDvT scaling corrections, further confirming the convergence of our results with increasing complexity of the CC approximations.

In terms of the Breit, QED, and basis extrapolation corrections to $\brad{nS_{1/2}}D\ketd{n'P_J}$, generally speaking, they become more and more important as $n'$ increases. For $\brad{6S_{1/2}}D\ketd{nP_{1/2}}$, they grow from a few 0.01\% for $n=6$ to a few percents for $n=12$, while for $\brad{6S_{1/2}}D\ketd{nP_{3/2}}$ and $\brad{7S_{1/2}}D\ketd{nP_{J}}$, the growth is less dramatic, from a few hundreths of a percent for $n=6$ to a few tenths of a percent for $n=12$. We also mention in passing that the relative roles of Breit and QED corrections in $\brad{nS_{1/2}}D\ketd{n'P_{1/2}}$ are noticeably more pronounced than those in $\brad{nS_{1/2}}D\ketd{n'P_{3/2}}$. The qualitative reason for this is due to the more relativistic character of the $p_{1/2}$ orbitals as compared to the $p_{3/2}$ orbitals. 

Although correlation effects on removal energies become less and less important with increasing principal quantum number, the same cannot be said, however, for all matrix elements. Indeed, Tables~\ref{tab:E1diople_6S1nP1} and~\ref{tab:E1diople_6S1nP3} show very large correlation corrections to $\brad{6S_{1/2}}D\ketd{nP_J}$ for $n\geq9$. Electron correlation, however, appears to have minimal effects on $\brad{7S_{1/2}}D\ketd{nP_J}$, as may be observed from Tables~\ref{tab:E1diople_7S1nP1} and~\ref{tab:E1diople_7S1nP3}. This may be qualitatively understood by noting that computing $\brad{nS_{1/2}}D\ketd{n'P_J}$ involves integrating products of wave functions which oscillate up to some point on the radial grid. Larger $n$ generally means more oscillations happening further away from the origin. A result is that if $n$ and $n'$ are very different, $\ket{nS_{1/2}}$ and $\ket{n'P_{J}}$ have disparate numbers of oscillations that happen at different places so their product oscillates for the whole integration range, thus yielding contributions that cancel instead of add to each other. This cancellation means that the integral depends delicately on the exact details of the wave functions, and small correlation corrections to the wave functions themselves could result in large corrections to the matrix elements. Other related features appear in Tables~\ref{tab:E1diople_6S1nP1} and~\ref{tab:E1diople_6S1nP3}: the RPA(DHF) approximation is particularly inadequate for $\brad{6S_{1/2}}D\ketd{10P_{1/2}}$ and $\brad{6S_{1/2}}D\ketd{11P_{1/2}}$ and the BO approximation seems to be doing poorly for all higher $n$. These artifacts are results of cancellations between the DHF and RPA contributions to the matrix elements, which become evident in detailed analyses of different contributions to the final CCSDpTvT results. 

Using the values of $\brad{nS_{1/2}}D\ketd{n'P_J}$, we computed the normalized ratio of reduced $E1$ matrix elements  $\xi_{nn'}$ connecting the $nS_{1/2}$ state to the two $n'P_J$ fine-structure states [see Eq.~\eqref{eq:effratio}]. The $\xi_{nn'}$ results are collected  in Tables~\ref{tab:ratio_6} and~\ref{tab:ratio_7}. The uncertainties in the final results for $\xi_{nn'}$ are also taken to be half the semiempirical scaling corrections. Note that we do not estimate the uncertainty for $\xi_{nn'}$ by adding the uncertainties for $\brad{nS_{1/2}}D\ketd{n'P_{1/2}}$ and $\brad{nS_{1/2}}D\ketd{n'P_{3/2}}$ in quadrature since they are not necessarily independent, given that the two matrix elements involve the same $nS_{1/2}$ state.

From Table~\ref{tab:ratio_7}, one observes that the ratio $\xi_{7,n}$ increases relatively slowly with increasing $n$, and that it remains quite close to the nonrelativistic value of unity. Table~\ref{tab:ratio_6} for $\xi_{6,n}$, on the other hand, tells a very different story. The ratio $\xi_{6,n}$ grows rapidly with increasing $n$, reaching $\xi_{6,12}\approx5.4$. This peculiarity may be understood by investigating the behaviors of the $\brad{6S_{1/2}}D\ketd{nP_{1/2}}$ and $\brad{6S_{1/2}}D\ketd{nP_{3/2}}$ matrix elements themselves. From Tables~\ref{tab:E1diople_6S1nP1} and~\ref{tab:E1diople_6S1nP3}, it appears that $\brad{6S_{1/2}}D\ketd{nP_{1/2}}$ is approaching zero as $n$ increases while $\brad{6S_{1/2}}D\ketd{nP_{3/2}}$ remains finite. This situation is similar to that of Cooper minima~\cite{FANO1968,Manson1985}, wherein the photoionization matrix element from the atomic ground state to the continuum $\varepsilon P_{1/2}$ state vanishes at a smaller continuum energy $\varepsilon$ than that to the continuum $\varepsilon P_{3/2}$ state. 

The previous comments on the various contributions to the matrix elements also apply to the ratio $\xi_{n,n'}$. In particular, the disparity in the Breit and QED corrections to the two $nP_J$ fine-structure components discussed above immediately translates into the ratios $\xi_{n,n'}$, whose relative Breit and QED corrections are similar to those of $\brad{nS_{1/2}}D\ketd{n'P_{1/2}}$. The spuriously large values for $\xi_{6,10}$ and $\xi_{6,11}$ in the RPA(DHF) approximation are due to the poor results from using the RPA(DHF) to estimate $\brad{6S_{1/2}}D\ketd{10P_{1/2}}$ and $\brad{6S_{1/2}}D\ketd{11P_{1/2}}$.

\begin{figure}
    \centering
    \includegraphics[width=\columnwidth]{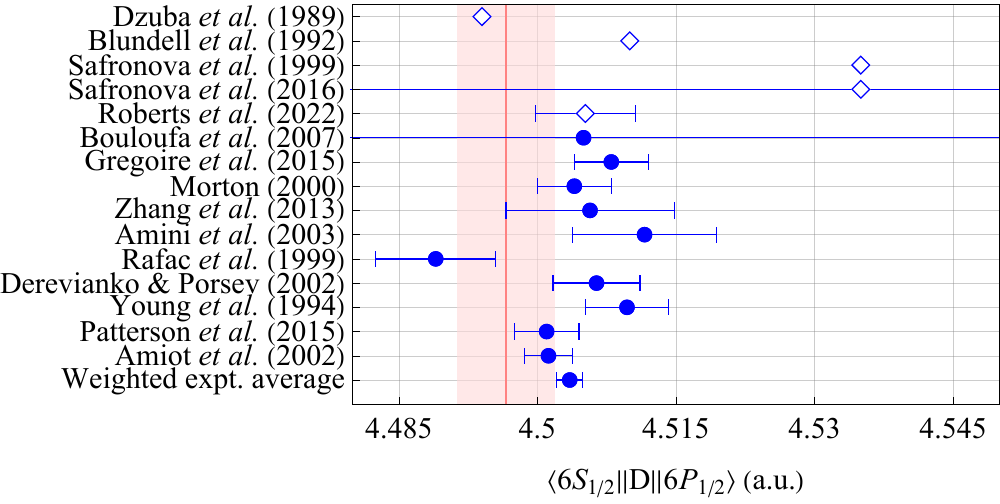}
    \caption{Comparison between our computed value (vertical line$+$uncertainty band) for the reduced electric-dipole matrix element $\brad{6S_{1/2}}D\ketd{6P_{1/2}}$  with existing experimental ($\bullet$) and theoretical ($\diamond$) results. The experimental results are ordered from the top down with decreasing uncertainties. The weighted average and uncertainty are computed using Eqs.~\eqref{eq:weightedave}.}
    \label{fig:mel.6s16p1}
\end{figure}

\begin{figure}
    \centering
    \includegraphics[width=\columnwidth]{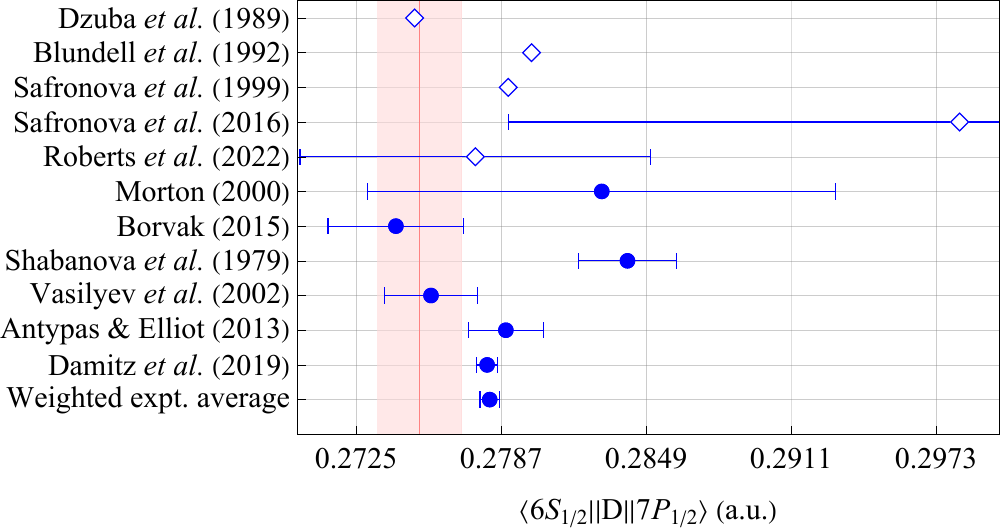}
    \caption{Comparison between our computed value (vertical line$+$uncertainty band) for the reduced electric-dipole matrix element $\brad{6S_{1/2}}D\ketd{7P_{1/2}}$  with existing experimental ($\bullet$) and theoretical ($\diamond$) results. The experimental results are ordered from the top down with decreasing uncertainties. The weighted average and uncertainty are computed using Eqs.~\eqref{eq:weightedave}.}
    \label{fig:mel.6s17p1}
\end{figure}

\begin{figure}
    \centering
    \includegraphics[width=\columnwidth]{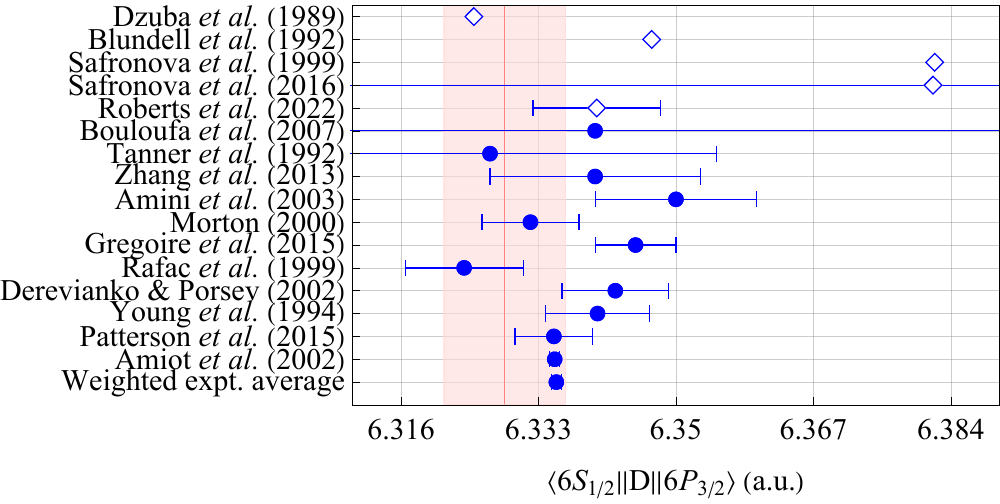}
    \caption{Comparison between our computed value (vertical line$+$uncertainty band) for the reduced electric-dipole matrix element $\brad{6S_{1/2}}D\ketd{6P_{3/2}}$  with existing experimental ($\bullet$) and theoretical ($\diamond$) results. The experimental results are ordered from the top down with decreasing uncertainties. The weighted average and uncertainty are computed using Eqs.~\eqref{eq:weightedave}.}
    \label{fig:mel.6s16p3}
\end{figure}

\begin{figure}
    \centering
    \includegraphics[width=\columnwidth]{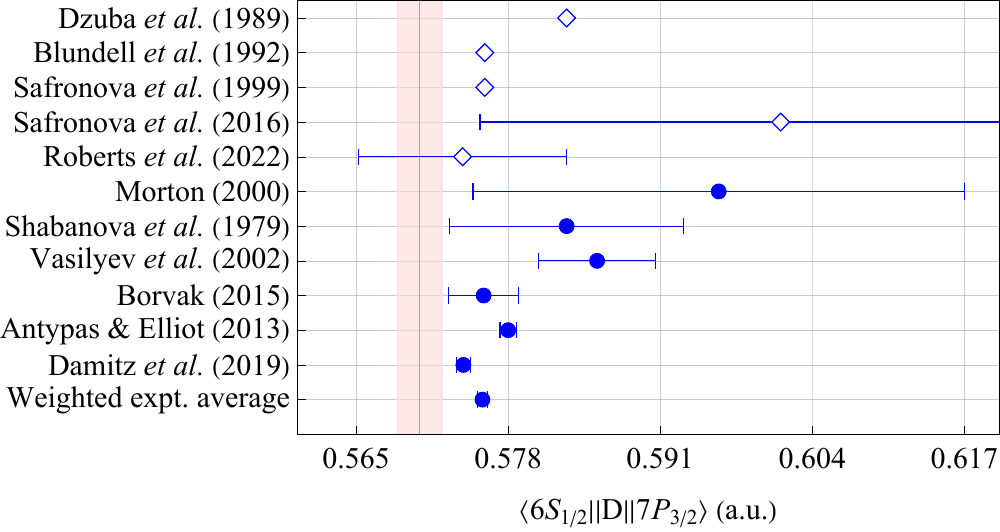}
    \caption{Comparison between our computed value (vertical line$+$uncertainty band) for the reduced electric-dipole matrix element $\brad{6S_{1/2}}D\ketd{7P_{3/2}}$  with existing experimental ($\bullet$) and theoretical ($\diamond$) results. The experimental results are ordered from the top down with decreasing uncertainties. The weighted average and uncertainty are computed using Eqs.~\eqref{eq:weightedave}.}
    \label{fig:mel.6s17p3}
\end{figure}

\begin{figure}
    \centering
    \includegraphics[width=\columnwidth]{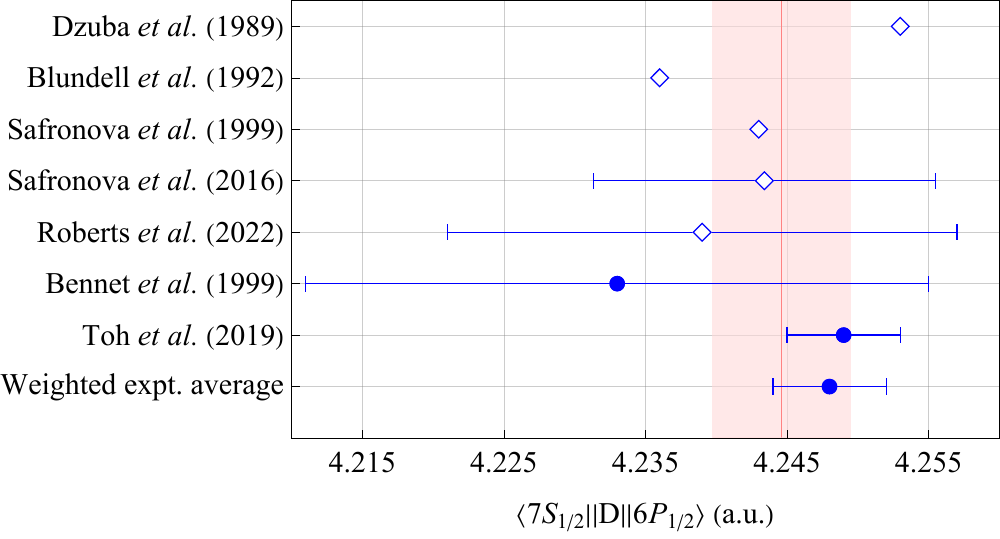}
    \caption{Comparison between our computed value (vertical line$+$uncertainty band) for the reduced electric-dipole matrix element $\brad{7S_{1/2}}D\ketd{6P_{1/2}}$  with existing experimental ($\bullet$) and theoretical ($\diamond$) results. The experimental results are ordered from the top down with decreasing uncertainties. The weighted average and uncertainty are computed using Eqs.~\eqref{eq:weightedave}.}
    \label{fig:mel.7s16p1}
\end{figure}

\begin{figure}
    \centering
    \includegraphics[width=\columnwidth]{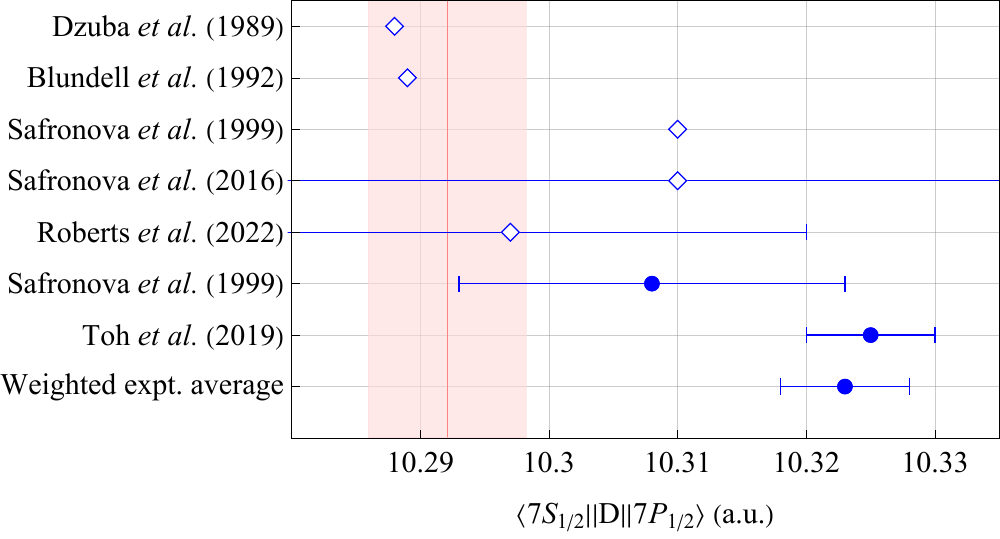}
    \caption{Comparison between our computed value (vertical line$+$uncertainty band) for the reduced electric-dipole matrix element $\brad{7S_{1/2}}D\ketd{7P_{1/2}}$  with existing experimental ($\bullet$) and theoretical ($\diamond$) results. The experimental results are ordered from the top down with decreasing uncertainties. The weighted average and uncertainty are computed using Eqs.~\eqref{eq:weightedave}.}
    \label{fig:mel.7s17p1}
\end{figure}

\begin{figure}
    \centering
    \includegraphics[width=\columnwidth]{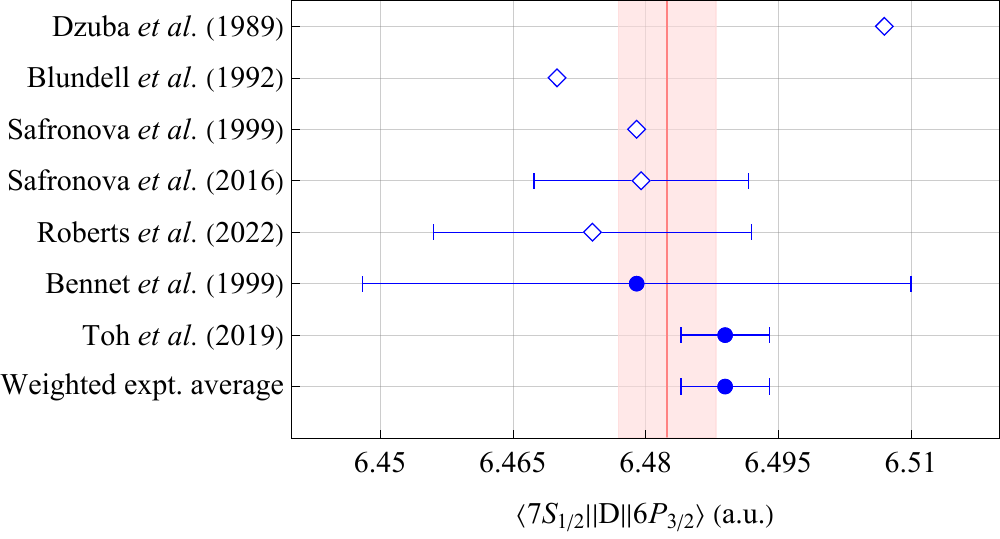}
    \caption{Comparison between our computed value (vertical line$+$uncertainty band) for the reduced electric-dipole matrix element $\brad{7S_{1/2}}D\ketd{6P_{3/2}}$  with existing experimental ($\bullet$) and theoretical ($\diamond$) results. The experimental results are ordered from the top down with decreasing uncertainties. The weighted average and uncertainty are computed using Eqs.~\eqref{eq:weightedave}.}
    \label{fig:mel.7s16p3}
\end{figure}

\begin{figure}
    \centering
    \includegraphics[width=\columnwidth]{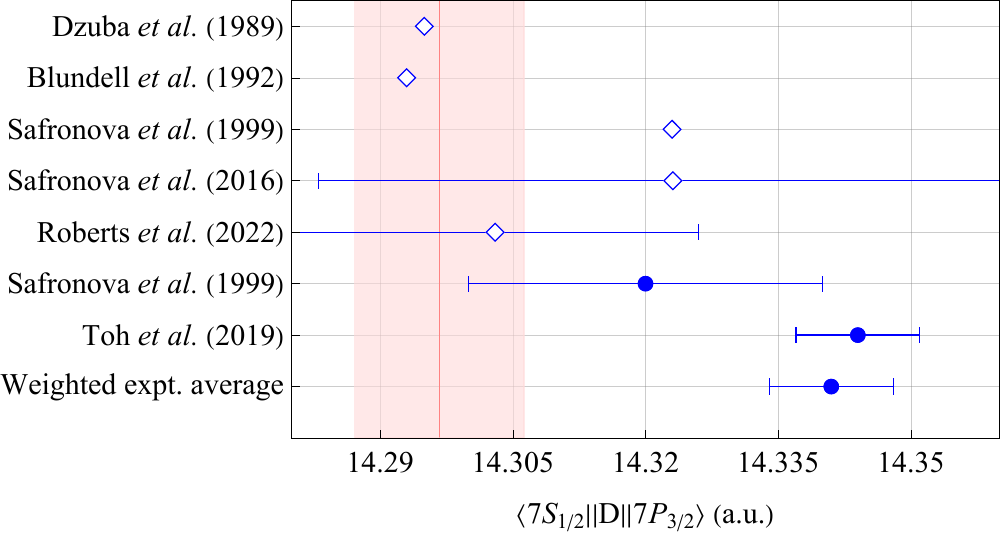}
    \caption{Comparison between our computed value (vertical line$+$uncertainty band) for the reduced electric-dipole matrix element $\brad{7S_{1/2}}D\ketd{7P_{3/2}}$  with existing experimental ($\bullet$) and theoretical ($\diamond$) results. The experimental results are ordered from the top down with decreasing uncertainties. The weighted average and uncertainty are computed using Eqs.~\eqref{eq:weightedave}.}
    \label{fig:mel.7s17p3}
\end{figure}

\begin{figure}
    \centering
    \includegraphics[width=\columnwidth]{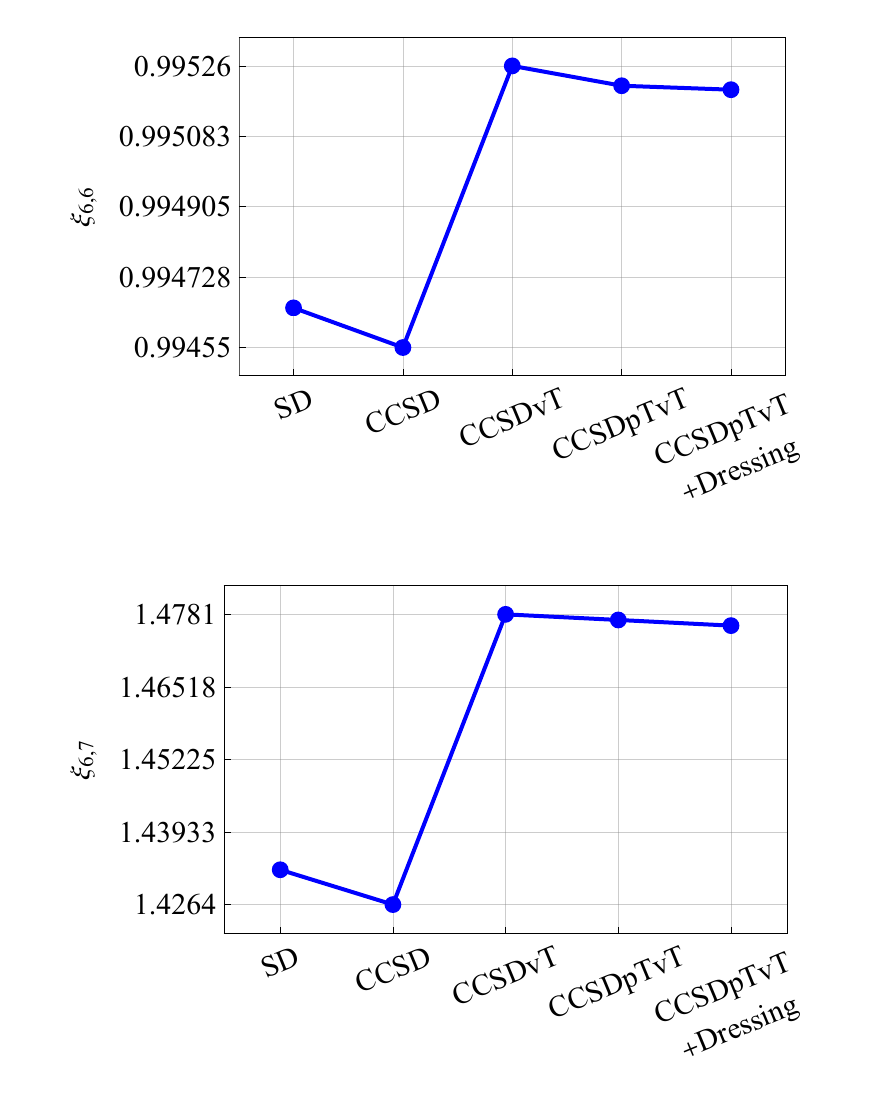}
    \caption{Convergence patterns for the normalized ratio $\xi_{6,n}\equiv(1/\sqrt{2})\brad{6S_{1/2}}D\ketd{nP_{3/2}}/\brad{6S_{1/2}}D\ketd{nP_{1/2}}$ with increasing complexity of the coupled-cluster method. The pattern for $n\geq8$ is similar to that of $n=7$.}\label{fig:convpattRat6s1}
\end{figure}

\begin{figure}
    \centering
    \includegraphics[width=\columnwidth]{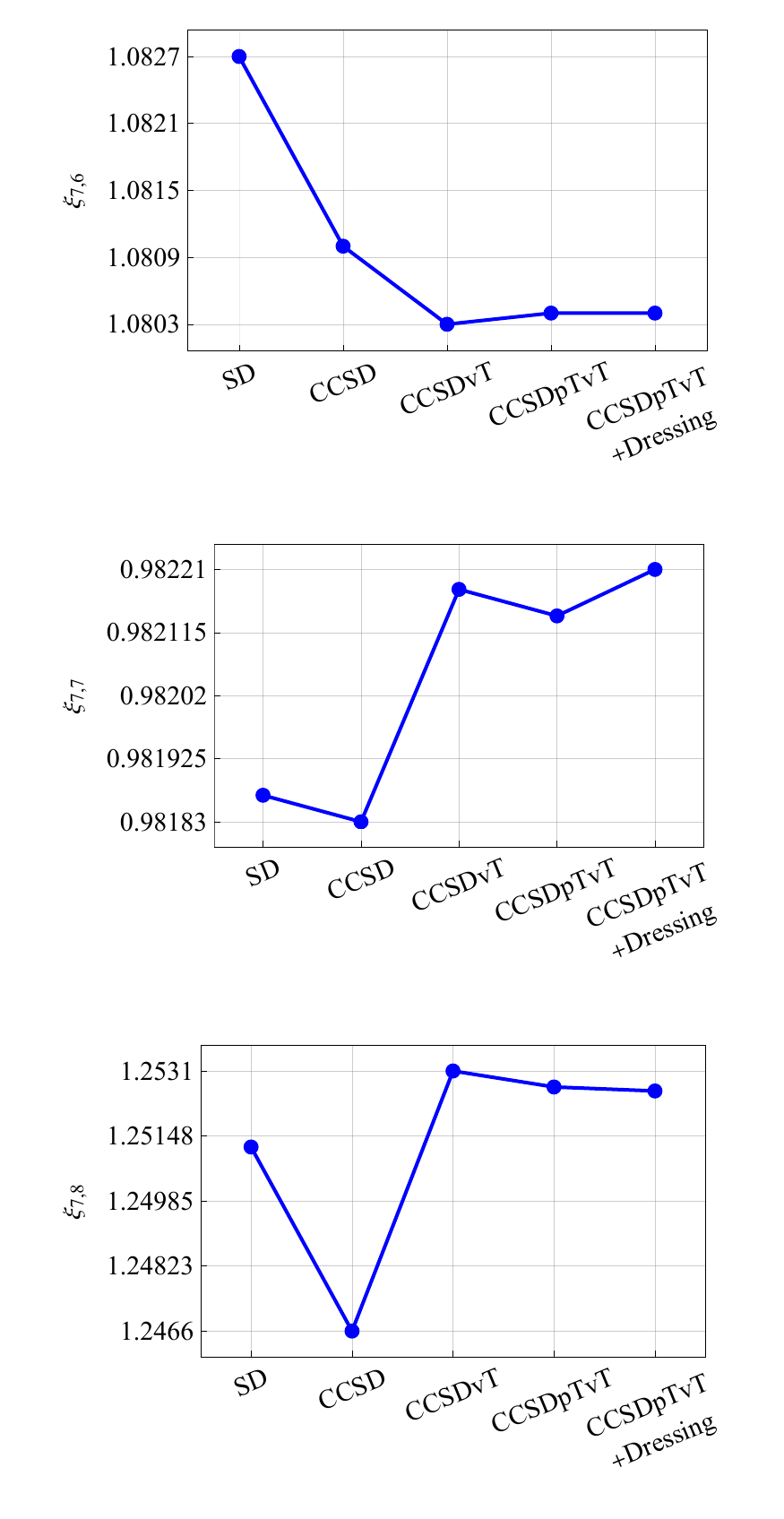}
    \caption{Convergence patterns for the normalized ratio $\xi_{7,n}\equiv(1/\sqrt{2})\brad{7S_{1/2}}D\ketd{nP_{3/2}}/\brad{7S_{1/2}}D\ketd{nP_{1/2}}$ with increasing complexity of the coupled-cluster method. The pattern for $n\geq9$ is similar to that for $n=8$.}\label{fig:convpattRat7s1}
\end{figure}

\begin{figure}
    \centering
    \includegraphics[width=\columnwidth]{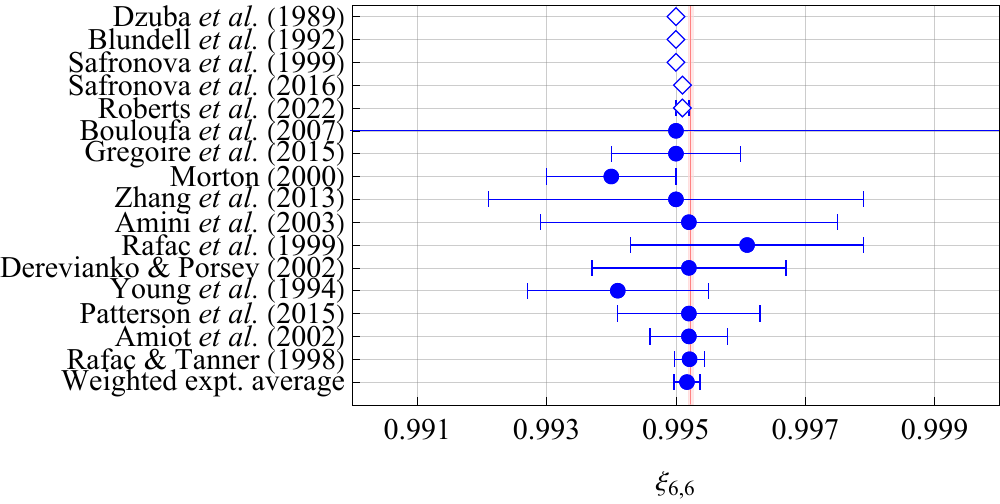}
    \caption{Comparison between our computed value (vertical line$+$uncertainty band) for the normalized ratio $\xi_{6,6}\equiv(1/\sqrt{2})\brad{6S_{1/2}}D\ketd{6P_{3/2}}/\brad{6S_{1/2}}D\ketd{6P_{1/2}}$  with existing experimental ($\bullet$) and theoretical ($\diamond$) results. The experimental results are ordered from the top down with decreasing uncertainties. The weighted average and uncertainty are computed using Eqs.~\eqref{eq:weightedave}.}
    \label{fig:rat66}
\end{figure}

\begin{figure}
    \centering
    \includegraphics[width=\columnwidth]{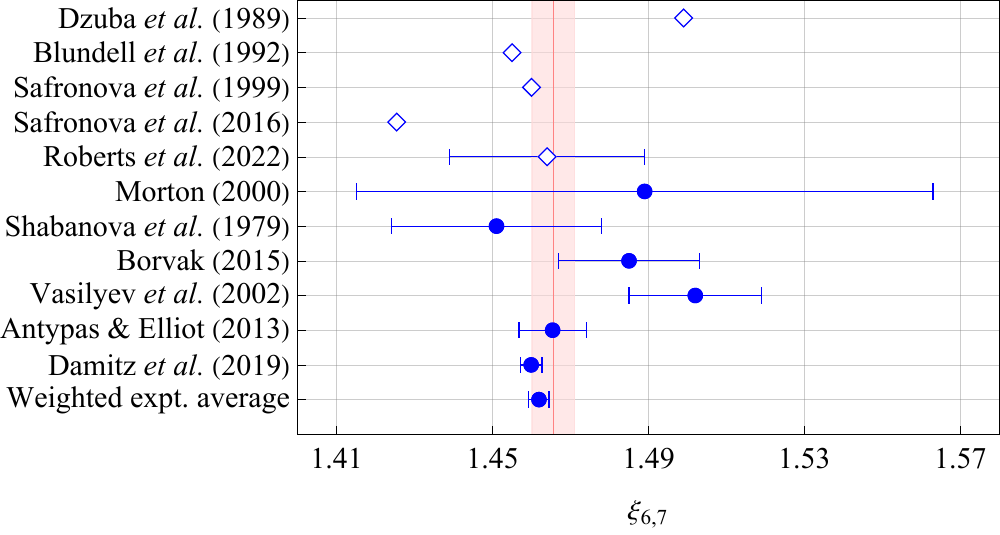}
    \caption{Comparison between our computed value (vertical line$+$uncertainty band) for the normalized ratio $\xi_{6,7}\equiv(1/\sqrt{2})\brad{6S_{1/2}}D\ketd{7P_{3/2}}/\brad{6S_{1/2}}D\ketd{7P_{1/2}}$  with existing experimental ($\bullet$) and theoretical ($\diamond$) results. The experimental results are ordered from the top down with decreasing uncertainties. The weighted average and uncertainty are computed using Eqs.~\eqref{eq:weightedave}.}
    \label{fig:rat67}
\end{figure}

\begin{figure}
	\centering
	\includegraphics[width=\columnwidth]{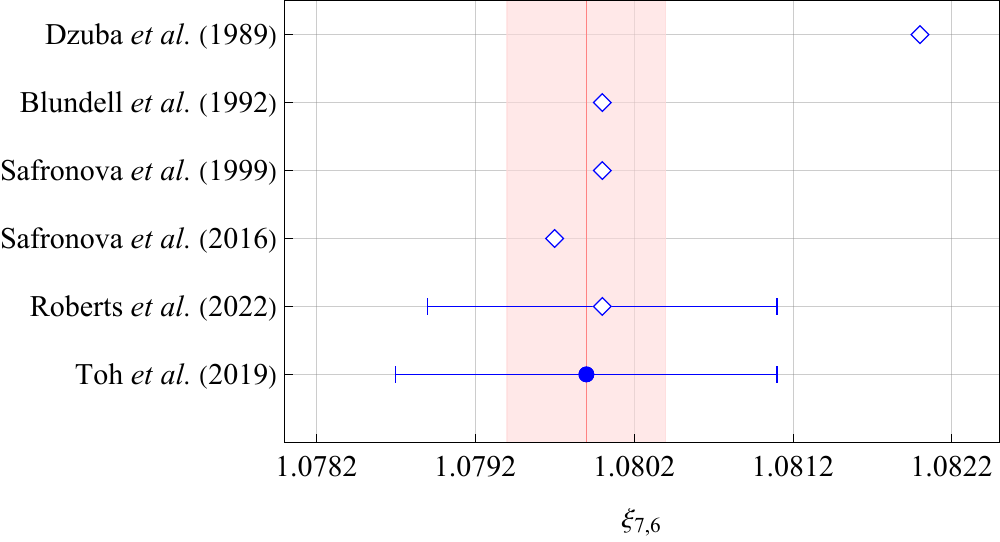}
	\caption{Comparison between our computed value (vertical line$+$uncertainty band) for the normalized ratio $\xi_{7,6}\equiv(1/\sqrt{2})\brad{7S_{1/2}}D\ketd{6P_{3/2}}/\brad{7S_{1/2}}D\ketd{6P_{1/2}}$  with existing experimental ($\bullet$) and theoretical ($\diamond$) results.}
	\label{fig:rat76}
\end{figure}

In Figs.~\ref{fig:mel.6s16p1}-\ref{fig:mel.7s17p3}, our computed values for the reduced $E1$ matrix elements are compared against existing experimental results as well as previous calculations. The convergence patterns for $\xi_{6,n}$ and $\xi_{7,n}$ with increasing complexity of the coupled-cluster approximation are shown in Figs.~\ref{fig:convpattRat6s1} and~\ref{fig:convpattRat7s1}. In Figs.~\ref{fig:rat66}-\ref{fig:rat76} our values for the normalized ratios $\xi_{6,6}$, $\xi_{6,7}$, and $\xi_{7,6}$ are compared against existing experimental results and previous calculations. The experimental weighted averages and uncertainties are computed using
\begin{subequations}\label{eq:weightedave}
    \begin{align}
        \bar{x}     &=\frac{\sum_ix_i/\sigma^2_i}{\sum_i1/\sigma^2_i}\,,\\
        \bar{\sigma}&=1/\sqrt{\sum_i1/\sigma^2_i}\,,
    \end{align}
\end{subequations}
where $x_i$ and $\sigma_i$ are the central value and uncertainty of each measurement. 

\begin{figure}
	\centering
	\includegraphics[width=\columnwidth]{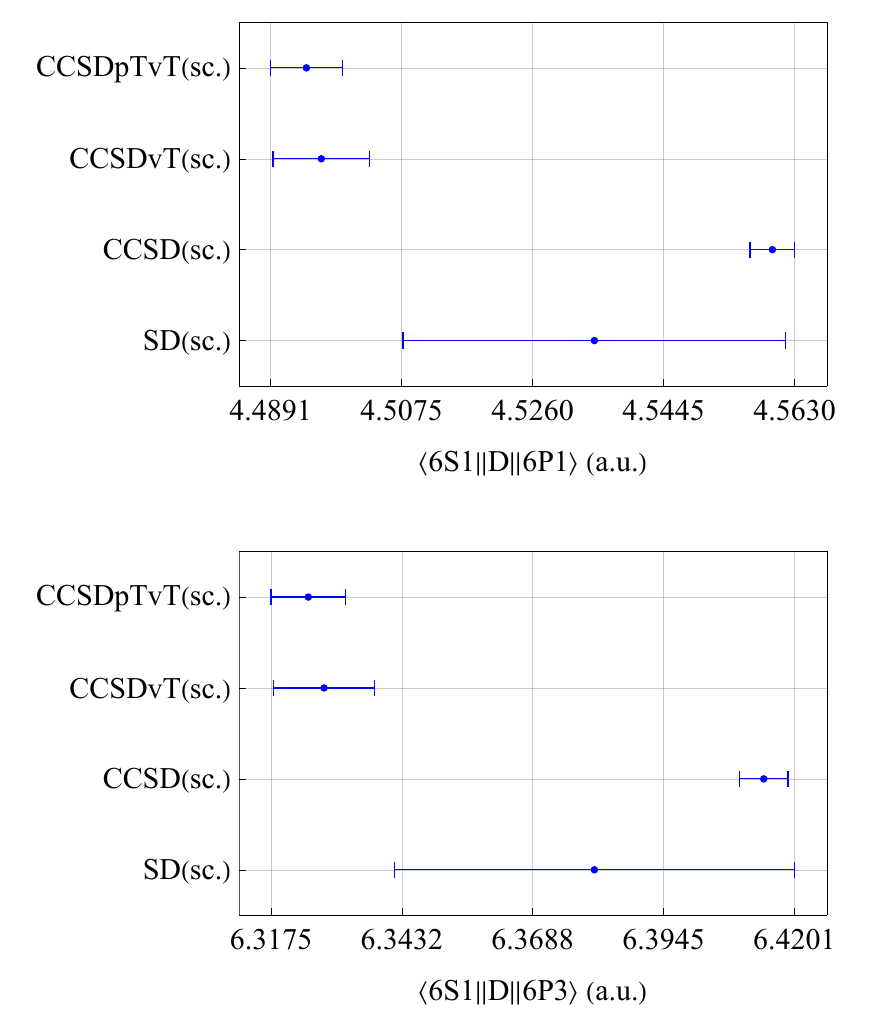}
	\caption{Comparison between the semiempirically scaled (sc.) $E1$ matrix elements at different levels of the coupled-cluster approximation. The error bars, representing the scaling uncertainties, are taken as half the difference between the scaled and unscaled values, i.e., $\left|\text{SD - SD(sc.)}\right|/2$ and so on.}
	\label{fig:CompareScaled} 
\end{figure}

When comparing our values with previous theoretical results, it is worth bearing in mind that the computations in Refs.~\cite{Blundell1992},~\cite{Safronova1999}, and \cite{Safronova2016} were performed at the SD and SDpT level, with semiempirical scaling included. A comparison of our SD results, both bare and with semiempirical scaling (not shown in Tables~\ref{tab:E1diople_6S1nP1}-\ref{tab:ratio_7}) and the values quoted in these earlier works shows excellent agreement. As a result, the differences between our results and earlier ones represent an improvement due to our accounting for higher-order terms in the CC approximation, most prominently nonlinear singles and doubles and valence triples. The improvement is noticeable in all cases and is significant for $\brad{6S_{1/2}}D\ketd{nP_{1/2}}$ with $n\geq9$. This also shows that the semiempirical scaling approach is only approximate and can only partially recover contributions from higher-order diagrams, as noted in Sec.~\ref{sec:Semiempirical scaling}. Indeed, although not shown in Tables~\ref{tab:E1diople_6S1nP1}-\ref{tab:ratio_7}, we also computed the scaled $E1$ matrix elements at the SD, CCSD, and CCSDvT levels. As shown in Fig.~\ref{fig:CompareScaled}, the scaled SD and scaled CCSD results are generally incompatible with the more complete scaled CCSDvT and scaled CCSDpTvT values.

Our results for $\brad{6,7S_{1/2}}D\ketd{6,7P_J}$ agree well with those of Ref.~\cite{Roberts2022}, which were obtained using the atomic many-body perturbation theory in the screened Coulomb interaction (AMPSCI), more colloquially known as the all-order Feynman technique. We remind the reader that AMPSCI involves summing to all orders perturbative series with respect to the screened Coulomb interaction, in contrast with the CC method, wherein the perturbative series are with respect to electron correlation. The Feynman technique thus misses certain diagrams with singles, doubles, and triples, but, on the other hand, includes some diagrams with quadruples not present in our CCSDpTvT calculations. We note that although earlier Feynman-technique values of Ref.~\cite{dzuba1989summation} for $\brad{6S_{1/2}}D\ketd{6,7P_{3/2}}$ and $\brad{7S_{1/2}}D\ketd{6P_J}$ disagree with ours, they also disagree with the more recent results of Ref.~\cite{Roberts2022}.

Overall, our results agree well with or are close to experimental data, except for $\brad{6S_{1/2}}D\ketd{12P_{1/2}}$ (16\% or 2.5$\sigma$ away), $\brad{6S_{1/2}}D\ketd{7P_{3/2}}$ (2.7$\sigma$ away), $\brad{7S_{1/2}}D\ketd{7P_{1/2}}$ (4.0$\sigma$ away) and $\brad{7S_{1/2}}D\ketd{7P_{3/2}}$(3.8$\sigma$ away). We point out, however, that with these disagreements, except for $\brad{6S_{1/2}}D\ketd{12P_{1/2}}$ which proves difficult due to strong cancellations making its value very small, the theory-experiment agreement is acceptable in terms of percentage. 

In relation to the determination of the APV amplitude in Cs, the relevant $E1$ matrix elements are those between $6,7S_{1/2}$ and $nP_{1/2}$ states. From Tables~\ref{tab:E1diople_6S1nP1} and~\ref{tab:E1diople_7S1nP1}, we observe that the main contributions, coming from $\brad{6S_{1/2}}D\ketd{6P_{1/2}}$, $\brad{7S_{1/2}}D\ketd{6P_{1/2}}$, and $\brad{7S_{1/2}}D\ketd{7P_{1/2}}$, have uncertainties $\sim0.1\%$. While other $E1$ matrix elements involving $P_{1/2}$ states with higher principle quantum numbers have larger uncertainties, their values are at least an order of magnitude smaller than those of the three main terms. As a result, the effective uncertainties arising from these ``tail'' terms are all sub-0.1\%. It is worth noting also that the largest uncertainty of 5.2\% in $\brad{6S_{1/2}}D\ketd{12P_{1/2}}$ is only half the uncertainty of the ``tail'' terms estimated in Ref.~\cite{Porsev2010}. As a result, although we do not claim that a determination of $E_{PV}({}^{133}{\rm Cs})$ using the $E1$ matrix elements quoted in this work will have a $\sim0.1\%$ uncertainty, such a level of accuracy is clearly reachable. Achieving this goal will be the subject of our future work based on a parity-mixed (PM) CC approach~\cite{BaoDiAndrei2022}, where the artificial separation of contributions to $E_{PV}$ into ``main'' and ``tail'' terms is circumvented. The results of the current paper will serve as gauges for the accuracy of the PM-CC approach. We note in passing that a new evaluation of $E_{PV}$ aiming at a $0.1\%$ uncertainty must also account for the contribution from neutrino vacuum polarization, which was recently estimated to be at the level of $\sim1\%$~\cite{Dzuba2022}.

We end this section with a few words on the computational cost associated with the different approximations employed. DHF, RPA, and BO are negligibly inexpensive. The SD computations take around $1/4$ core-hour for $S_{1/2}$ and $P_{1/2}$ states and around $1$ core-hour for $P_{3/2}$ states. CCSD computations cost around $2.5$ core-hours for $S_{1/2}$ and $P_{1/2}$ states and around $5$ core-hour for $P_{3/2}$ states. Calculations involving valence triple excitations are quite expensive: on our computer server with 160 cores, $S_{1/2}$ and $P_{1/2}$ states take around 8 real-time hours per state and $P_{3/2}$ states take around 22 real-time hours per state. The inclusion of perturbative core triples does not drastically increase the computational cost compared to CCSD and CCSDvT.

\begin{table*}[ht]
    \centering
    \begin{tabular}{ld{7}d{10}d{10}d{9}d{9}d{9}d{10}}
    \hline\hline
    \multicolumn{1}{c}{$6S_{1/2}\rightarrow$}& \multicolumn{1}{c}{$6P_{1/2}$}&\multicolumn{1}{c}{$7P_{1/2}$}&\multicolumn{1}{c}{$8P_{1/2}$} &\multicolumn{1}{c}{$9P_{1/2}$}&\multicolumn{1}{c}{$10P_{1/2}$}& \multicolumn{1}{c}{$11P_{1/2}$}& \multicolumn{1}{c}{$12P_{1/2}$}\\
    \hline
    DHF                    & 5.2777 & 3.7174[-1] & 1.3262[-1] & 7.1742[-2]       & 4.6735[-2] & 3.3731[-2] & 2.5952[-2] \\
    RPA(DHF)              & 4.9744 & 2.3872[-1] & 0.4983[-1] & 1.3121[-2]       & 0.2197[-2] & 0.1679[-2] & 0.3118[-2] \\
    BO                     & 4.7250 & 4.4414[-1] & 1.8142[-1] & 10.601\,\,\,[-2] & 7.2457[-2] & 5.3975[-2] & 4.2430[-2] \\
    RPA(BO)               & 4.3909 & 3.0269[-1] & 0.9402[-1] & 4.4344[-2]       & 2.5708[-2] & 1.6871[-2] & 1.2019[-2] \\
    SD                     & 4.4806 & 2.9655[-1] & 0.9060[-1] & 4.2257[-2]       & 2.4291[-2] & 1.5841[-2] & 1.1240[-2] \\
    CCSD                   & 4.5535 & 3.0274[-1] & 0.9285[-1] & 4.3478[-2]       & 2.5079[-2] & 1.6395[-2] & 1.1645[-2] \\
    CCSDpT                 & 4.5480 & 3.0299[-1] & 0.9301[-1] & 4.3587[-2]       & 2.5157[-2] & 1.6455[-2] & 1.1693[-2] \\
    CCSDvT                 & 4.5098 & 2.7138[-1] & 0.7314[-1] & 2.9477[-2]       & 1.4421[-2] & 0.7912[-2] & 0.4678[-2] \\
    CCSDpTvT               & 4.5042 & 2.7163[-1] & 0.7330[-1] & 2.9583[-2]       & 1.4498[-2] & 0.7971[-2] & 0.4725[-2] \\
    \hline     
    \multicolumn{8}{c}{Other corrections}\\
    Scaling                &-0.0101	& 0.0280[-1] & 0.0155[-1] & 0.0858[-2]       & 0.0547[-2] & 0.0580[-2] & 0.0361[-2]  \\

    Dressing               & 0.0017 & 0.0065[-1] & 0.0040[-1] & 0.0277[-2]       & 0.0210[-2] & 0.0167[-2] & 0.0137[-2] \\
    Breit                  &-0.0010 & 0.0189[-1] & 0.0107[-1] & 0.0712[-2]       & 0.0523[-2] & 0.0407[-2] & 0.0328[-2] \\
    QED                    & 0.0035 &-0.0225[-1] &-0.0131[-1] &-0.0882[-2]       &-0.0652[-2] &-0.0509[-2] &-0.0413[-2] \\
    Basis extrapolation    &-0.0017 & 0.0050[-1] & 0.0032[-1] & 0.0221[-2]       & 0.0165[-2] & 0.0130[-2] & 0.0106[-2] \\
    \hline
    Final result           & 4.4966(52) & 2.752(18)[-1] & 0.753(10)[-1] & 3.077(61)[-2] & 1.529(42)[-2] & 0.875(38)[-2] & 0.524(27)[-2] \\
    Uncertainty (\%)       & 0.12 & 0.65 & 1.3 & 2.0 & 2.7 & 4.4 & 5.2 \\
    \hline
    Other results          & 4.5052(54)\footnote{\label{Roberts2022-1}Roberts \textit{et al.}~(2022),~Ref.~\cite{Roberts2022}} & 2.776(75)[-1]\footnotemark[1] &&&&&\\ 
    					  & 4.535(77)\footnote{\label{Safronova2016-1}Safronova \textit{et al.}~(2016),~Ref.~\cite{Safronova2016}} & 2.98(19)[-1]\footnotemark[2] & 0.92(10)[-1]\footnotemark[2] & 4.29(68)[-2]\footnotemark[2] & 2.48(50)[-2]\footnotemark[2] & 1.62(39)[-2]\footnotemark[2] & 1.15(32)[-2]\footnotemark[2] \\
                           & 4.535\footnote{\label{Safronova1999-1}Safronova \textit{et al.}~(1999),~Ref.~\cite{Safronova1999}} & 2.79[-1]\footnotemark[3] & 0.81[-1]\footnotemark[3] &&&&\\
                           & 4.510\footnote{\label{Blundell1992-1}Blundell \textit{et al.}~(1992),~Ref.~\cite{Blundell1992}} & 2.80[-1]\footnotemark[4] & 0.78[-1]\footnotemark[4] &&&&\\
                           & 4.494\footnote{\label{Dzuba1989-1}Dzuba \textit{et al.}~(1989),~Ref.~\cite{dzuba1989summation}} & 2.75[-1]\footnotemark[5] &&&&&\\
    \hline
    Experiments            & 4.5012(26)\footnote{Amiot \textit{et al.}~(2002),~Ref.~\cite{Amiot2002}} & 2.7810(45)[-1]\footnote{Damitz \textit{et al.}~(2019),~Ref.~\cite{Damitz2019}} & 0.723(44)[-1]\footnotemark[18] & 3.23(37)[-2]\footnotemark[18] & 1.62(8)[-2]\footnotemark[18] & 0.957(46)[-2]\footnotemark[18] & 0.627(30)[-2]\footnotemark[18] \\
                           & 4.5010(35)\footnote{Patterson \textit{et al.}~(2015),~Ref.~\cite{Patterson2015}} & 2.789(16)[-1]\footnote{Antypas \& Elliot (2013),~Ref.~\cite{Antypas2013}} &&&&&\\
                           & 4.5097(45)\footnote{Young \textit{et al.}~(1994),~Ref.~\cite{Young1994}} & 2.757(20)[-1]\footnote{Vasilyev \textit{et al.}~(2002),~Ref.~\cite{Vasilyev2002}} &&&&&\\
                           & 4.5064(47)\footnote{Derevianko \& Porsev~(2002),~Ref.~\cite{Derevianko2002}} & 2.841(21)[-1]\footnote{Shabanova \textit{et al.}~(1979),~Ref.~\cite{ShabMonKhl79}} &&&&&\\
                           & 4.4890(65)\footnote{Rafac \textit{et al.}~(1999),~Ref.~\cite{Rafac1999}} & 2.742(29)[-1]\footnote{Borvak (2015),~Ref.~\cite{borvak2015direct}} &&&&&\\
                           & 4.5116(78)\footnote{Amini \textit{et al.}~(2003),~Ref.~\cite{Amini2003}} & 2.83(10)[-1]\footnotemark[18] &&&&&\\
                           & 4.5057(91)\footnote{Zhang \textit{et al.}~(2013),~Ref.~\cite{Zhang2013}} &&&&&&\\
                           & 4.504(4)\footnote{\label{00MOR}Morton~(2000),~Ref.~\cite{Morton_2000}} &&&&&&\\
                           & 4.508(4)\footnote{Gregoire \textit{et al.}~(2015),~Ref.~\cite{Gregoire2015}} &&&&&&\\
                           & 4.505(47)\footnote{Bouloufa \textit{et al.}~(2007),~Ref.~\cite{Bouloufa2007}} &&&&&&\\
    Weighted average       & 4.5035(14) & 2.7822(42)[-1]                                                       &&&&&\\
    \hline
    Difference (\%)         & -0.15 & -1.1 & 4.2  & -4.7  & -5.6 & -8.6 & -16 \\
    Difference ($\sigma$)   & -1.3  & -1.6 & 0.67 & -0.41 & -1.0 & -1.4 & -2.5 \\
    \hline\hline
    \end{tabular}
    \caption{Reduced electric-dipole matrix elements $\brad{6S_{1/2}}D\ketd{n'P_{1/2}}$ (in atomic units a.u.) for $n'=6-12$ in Cs with various approximations. See Table~\ref{tab:Energy_S1} caption for explanation of entries. The notation $x[y]$ stands for $x\times 10^y$.
    }\label{tab:E1diople_6S1nP1}
\end{table*}

\begin{table*}[ht]
    \centering
    \begin{tabular}{ld{7}d{11}d{10}d{10}d{9}d{9}d{10}}
    \hline\hline
    \multicolumn{1}{c}{$6S_{1/2}\rightarrow$}& \multicolumn{1}{c}{$6P_{3/2}$}&\multicolumn{1}{c}{$7P_{3/2}$}&\multicolumn{1}{c}{$8P_{3/2}$} &\multicolumn{1}{c}{$9P_{3/2}$}&\multicolumn{1}{c}{$10P_{3/2}$}& \multicolumn{1}{c}{$11P_{3/2}$}& \multicolumn{1}{c}{$12P_{3/2}$}\\
    \hline
    DHF                    & 7.4264 & 6.9474[-1] & 2.8323[-1] & 1.6582[-1] & 11.359\,\,\,[-2] & 8.4797[-2]       & 6.6791[-2] \\
    RPA(DHF)              & 7.0131 & 5.0875[-1] & 1.6648[-1] & 0.8280[-1] & 5.0362[-2]       & 3.4443[-2]       & 2.5408[-2] \\
    BO                     & 6.6251 & 8.0698[-1] & 3.6037[-1] & 2.2047[-1] & 15.480\,\,\,[-2] & 11.731\,\,\,[-2] & 9.3300[-2] \\
    RPA(BO)               & 6.1740 & 6.0914[-1] & 2.3686[-1] & 1.3289[-1] & 8.8212[-2]       & 6.4364[-2]       & 4.9843[-2] \\
    SD                     & 6.3026 & 6.0083[-1] & 2.3174[-1] & 1.2960[-1] & 8.5908[-2]       & 6.2645[-2]       & 4.8510[-2] \\
    CCSD                   & 6.4045 & 6.1071[-1] & 2.3565[-1] & 1.3185[-1] & 8.7425[-2]       & 6.3754[-2]       & 4.9358[-2] \\
    CCSDpT                 & 6.3966 & 6.1098[-1] & 2.3580[-1] & 1.3195[-1] & 8.7495[-2]       & 6.3807[-2]       & 4.9400[-2] \\
    CCSDvT                 & 6.3476 & 5.6727[-1] & 2.0802[-1] & 1.1212[-1] & 7.2360[-2]       & 5.1742[-2]       & 3.9480[-2] \\
    CCSDpTvT               & 6.3394 & 5.6740[-1] & 2.0814[-1] & 1.1220[-1] & 7.2419[-2]       & 5.1786[-2]       & 3.9516[-2] \\
    \hline     
    \multicolumn{8}{c}{Other corrections}\\
    Scaling                &-0.0146 & 0.0355[-1] & 0.0184[-1] & 0.013[-1] & 0.0686[-2]        & 0.0871[-2]       & 0.0702[-2] \\
    Dressing               & 0.0023 & 0.0096[-1] & 0.0059[-1] & 0.0042[-1] & 0.0319[-2]       & 0.0254[-2]       & 0.0209[-2] \\
    Breit                  &-0.0011 & 0.0051[-1] & 0.0029[-1] & 0.0019[-1] & 0.0138[-2]       & 0.0107[-2]       & 0.0086[-2] \\
    QED                    & 0.0052 &-0.0251[-1] &-0.0152[-1] &-0.0104[-1] &-0.0776[-2]       &-0.0609[-2]       &-0.0495[-2] \\
    Basis extrapolation    &-0.0024 & 0.0052[-1] & 0.0035[-1] & 0.0025[-1] & 0.0188[-2]       & 0.0149[-2]       & 0.0122[-2] \\
    \hline
    Final result           & 6.3288(75) & 5.704(19)[-1] & 2.097(10)[-1] & 1.1332(72)[-1] & 7.297(41)[-2] & 5.256(47)[-2] & 4.014(38)[-2] \\
    Uncertainty (\%)       & 0.12 & 0.34 & 0.49 & 0.63 & 0.56 & 0.89 & 0.95\\
    \hline
    Other resuts           & 6.3402(79) \footnote{\label{Roberts2022-2}Roberts \textit{et al.}~(2022),~Ref.~\cite{Roberts2022}} & 5.741(89) [-1]\footnotemark[1] &&&&&\\ 
    					   & 6.382(79)\footnote{\label{Safronova2016-2}Safronova \textit{et al.}~(2016),~Ref.~\cite{Safronova2016}} & 6.01(26)[-1]\footnotemark[2] & 2.32(14)[-1]\footnotemark[2] & 1.297(96)[-1]\footnotemark[2] & 8.60(71)[-2]\footnotemark[2] & 6.27(56)[-2]\footnotemark[2] & 4.86(46)[-2]\footnotemark[2]\\
                           & 6.382\footnote{\label{Safronova1999-2}Safronova \textit{et al.}~(1999),~Ref.~\cite{Safronova1999}} & 5.76[-1]\footnotemark[3] & 2.18[-1]\footnotemark[3] &&&&\\
                           & 6.347\footnote{\label{Blundell1992-2}Blundell \textit{et al.}~(1992),~Ref.~\cite{Blundell1992}} & 5.76[-1]\footnotemark[4] & 2.14[-1]\footnotemark[4] &&&&\\
                           & 6.325\footnote{\label{Dzuba1989-2}Dzuba \textit{et al.}~(1989),~Ref.~\cite{dzuba1989summation}} & 5.83[-1]\footnotemark[5] &&&&&\\
    \hline
    Experiment             & 6.3350(6)\footnote{Amiot \textit{et al.}~(2002),~Ref.~\cite{Amiot2002}} & 5.7417(57)[-1]\footnote{Damitz \textit{et al.}~(2019),~Ref.~\cite{Damitz2019}} & 2.11(8)[-1]\footnotemark[17] & 1.15(7)[-1]\footnotemark[17] & 7.22(34)[-2]\footnotemark[17] & 5.29(46)[-2]\footnotemark[17] & 3.98(19)[-2]\footnotemark[17] \\
                           & 6.3349(48)\footnote{Patterson \textit{et al.}~(2015),~Ref.~\cite{Patterson2015}} & 5.780(7)[-1]\footnote{Antypas \& Elliot (2013),~Ref.~\cite{Antypas2013}} &&&&&\\
                           & 6.3403(64)\footnote{Young \textit{et al.}~(1994),~Ref.~\cite{Young1994}} & 5.759(30)[-1]\footnote{Borvak (2015),~Ref.~\cite{borvak2015direct}} &&&&&\\
                           & 6.3425(66)\footnote{Derevianko \& Porsev~(2002),~Ref.~\cite{Derevianko2002}} & 5.856(50)[-1]\footnote{Vasilyev \textit{et al.}~(2002),~Ref.~\cite{Vasilyev2002}} &&&&&\\
                           & 6.3238(73)\footnote{Rafac \textit{et al.}~(1999),~Ref.~\cite{Rafac1999}} & 5.83(10)[-1]\footnote{Shabanova \textit{et al.}~(1979),~Ref.~\cite{ShabMonKhl79}} &&&&&\\
                           & 6.345(5)\footnote{Gregoire \textit{et al.}~(2015),~Ref.~\cite{Gregoire2015}} & 5.96(21)[-1]\footnotemark[17] &&&&&\\
                           & 6.332(6)\footnote{\label{00MOR2}Morton~(2000),~Ref.~\cite{Morton_2000}} &&&&&&\\
                           & 6.350(10)\footnote{Amini \textit{et al.}~(2003),~Ref.~\cite{Amini2003}} &&&&&&\\
                           & 6.340(13)\footnote{Zhang \textit{et al.}~(2013),~Ref.~\cite{Zhang2013}} &&&&&&\\
                           & 6.327(28)\footnote{Tanner \textit{et al.}~(1992),~Ref.~\cite{Tanner1992}} &&&&&&\\
                           & 6.340(63)\footnote{Bouloufa \textit{et al.}~(2007),~Ref.~\cite{Bouloufa2007}} &&&&&&\\
    Weighted average       & 6.3352(6) & 5.7580(44)[-1] &&&&&\\
    \hline
    Difference (\%)         & -0.10 & -0.93 & -0.62  & -1.5  & 1.1  & -0.65 & 0.85  \\
    Difference ($\sigma$)   & -0.8  & -2.7  & -0.16  & -0.24 & 0.23 & -0.07 & 0.18 \\
    \hline\hline
    \end{tabular}
    \caption{Reduced electric-dipole matrix elements $\brad{6S_{1/2}}D\ketd{n'P_{3/2}}$ (in atomic units a.u.) for $n'=6-12$ in Cs with various approximations. See Table~\ref{tab:Energy_S1} caption for explanation of entries. The notation $x[y]$ stands for $x\times 10^y$.}\label{tab:E1diople_6S1nP3}
\end{table*}

\begin{table*}[ht]
    \centering
    \begin{tabular}{ld{8}d{11}d{10}d{10}d{10}d{9}d{10}}
    \hline\hline
    \multicolumn{1}{c}{$7S_{1/2}\rightarrow$}& \multicolumn{1}{c}{$6P_{1/2}$}&\multicolumn{1}{c}{$7P_{1/2}$}&\multicolumn{1}{c}{$8P_{1/2}$} &\multicolumn{1}{c}{$9P_{1/2}$}&\multicolumn{1}{c}{$10P_{1/2}$}& \multicolumn{1}{c}{$11P_{1/2}$}& \multicolumn{1}{c}{$12P_{1/2}$}\\
    \hline
    DHF                    & 4.4131 & 1.1009[1] & 9.2117[-1]              & 3.3720[-1] & 1.8332[-1] & 1.1944[-1] & 8.6149[-2] \\
    RPA(DHF)              & 4.4494 & 1.0921[1] & 8.6912[-1]              & 3.0076[-1] & 1.5576[-1] & 0.9758[-1] & 6.8225[-2] \\
    BO                     & 4.1945 & 1.0263[1] & 10.080\,\,\,[-1]        & 3.9870[-1] & 2.2756[-1] & 1.5320[-1] & 11.303\,\,\,[-2] \\
    RPA(BO)               & 4.2232 & 1.0175[1] & 9.5622[-1]              & 3.6260[-1] & 2.0034[-1] & 1.3166[-1] & 9.5406[-2] \\
    SD                     & 4.1952 & 1.0253[1] & 9.2901[-1]              & 3.4658[-1] & 1.8956[-1] & 1.2374[-1] & 8.9261[-2] \\
    CCSD                   & 4.2502 & 1.0298[1] & 9.4069[-1]              & 3.5176[-1] & 1.9263[-1] & 1.2582[-1] & 9.0782[-2] \\
    CCSDpT                 & 4.2497 & 1.0292[1] & 9.4155[-1]              & 3.5228[-1] & 1.9299[-1] & 1.2608[-1] & 9.0988[-2] \\
    CCSDvT                 & 4.2527 & 1.0308[1] & 9.2122[-1]              & 3.3906[-1] & 1.8342[-1] & 1.1870[-1] & 8.5035[-2] \\
    CCSDpTvT               & 4.2522 & 1.0302[1] & 9.2210[-1]              & 3.3960[-1] & 1.8379[-1] & 1.1897[-1] & 8.5246[-2] \\
    \hline     
    \multicolumn{8}{c}{Other corrections}\\  
    Scaling                &-0.0081 &-0.0011[1] & 0.0383[-1]              & 0.0150[-1] & 0.0078[-1] & 0.0090[-1] & 0.0471[-2] \\
    Dressing               & 0.0000 & 0.0000[1] & 0.0033[-1]              & 0.0023[-1] & 0.0017[-1] & 0.0014[-1] & 0.0111[-2] \\
    Breit                  & 0.0049 &-0.0003[1] & 0.0342[-1]              & 0.0195[-1] & 0.0131[-1] & 0.0096[-1] & 0.0754[-2] \\
    QED                    &-0.0045 & 0.0007[1] &-0.0423[-1]              &-0.0244[-1] &-0.0165[-1] &-0.0122[-1] &-0.0957[-2] \\
    Basis extrapolation    & 0.0001 &-0.0003[1] & 0.0083[-1]              & 0.0051[-1] & 0.0035[-1] & 0.0026[-1] & 0.0207[-2] \\
    \hline
    Final result           & 4.2446(49) & 1.0292(6)[1] & 9.263(28)[-1] & 3.414(14)[-1] & 1.8475(88)[-1] & 1.2002(74)[-1] & 8.583(52)[-2] \\
    Uncertainty (\%)       & 0.1 & 0.060 & 0.30 & 0.41 & 0.48 & 0.62 & 0.60\\
    \hline
    Other results          & 4.239(18) \footnote{\label{Roberts2022-3}Roberts \textit{et al.}~(2022),~Ref.~\cite{Roberts2022}} & 1.0297(23)[1]\footnotemark[1] &&&&&\\ 
    					   & 4.243(12)\footnote{\label{Safronova2016-3}Safronova \textit{et al.}~(2016),~Ref.~\cite{Safronova2016}} & 1.0310(40)[1]\footnotemark[2] & 9.14(27)[-1]\footnotemark[2] & 3.49(10)\footnotemark[2] & 1.908(60)[-1]\footnotemark[2] & 1.247(44)[-1]\footnotemark[2] & 9.00(35)[-2]\footnotemark[2]\\
                           & 4.243\footnote{\label{Safronova1999-3}Safronova \textit{et al.}~(1999),~Ref.~\cite{Safronova1999}} & 1.0310[1]\footnotemark[3] &&&&&\\
                           & 4.236\footnote{\label{Blundell1992-3}Blundell \textit{et al.}~(1992),~Ref.~\cite{Blundell1992}} & 1.0289[1]\footnotemark[4] &&&&&\\
                           & 4.253\footnote{\label{Dzuba1989-3}Dzuba \textit{et al.}~(1989),~Ref.~\cite{dzuba1989summation}} & 1.0288[1]\footnotemark[5] &&&&&\\
    \hline
    Experiment             & 4.249(4)\footnote{\label{Toh2019Mels}Toh \textit{et al.}~(2019),~Ref.~\cite{Toh2019Mels}}       & 1.0325(5)[1]\footnote{\label{Toh2019Polarizabilities}Toh \textit{et al.}~(2019),~Ref.~\cite{Toh2019Polarizabilities}} &&&&&\\
                           & 4.233(22)\footnote{Bouchiat \textit{et al.}~(1984),~Ref.~\cite{Bouchiat1984}} & 1.0308(15)[1]\footnote{Bennett \textit{et al.}~(1999),~Ref.~\cite{PhysRevA.59.R16}} &&&&&\\
    Weighted average       & 4.248(4) & 1.0323(5)[1]                                                                 &&&&&\\
    \hline
    Difference (\%)         & -0.09 & -0.30 &&&&&\\
    Difference ($\sigma$)   & -0.62  & -4.0 &&&&&\\
    \hline\hline
    \end{tabular}
    \caption{Reduced electric-dipole matrix elements $\brad{7S_{1/2}}D\ketd{n'P_{1/2}}$ (in atomic units a.u.) for $n'=6-12$ in Cs with various approximations. See Table~\ref{tab:Energy_S1} caption for explanation of entries. The notation $x[y]$ stands for $x\times 10^y$.}\label{tab:E1diople_7S1nP1}
\end{table*}

\begin{table*}[ht]
    \centering
    \begin{tabular}{ld{7}d{9}d{7}d{10}d{10}d{10}d{10}}
    \hline\hline
    \multicolumn{1}{c}{$7S_{1/2}\rightarrow$}& 
    \multicolumn{1}{c}{$6P_{3/2}$}&
    \multicolumn{1}{c}{$7P_{3/2}$}&
    \multicolumn{1}{c}{$8P_{3/2}$} &
    \multicolumn{1}{c}{$9P_{3/2}$}&
    \multicolumn{1}{c}{$10P_{3/2}$}& 
    \multicolumn{1}{c}{$11P_{3/2}$}& 
    \multicolumn{1}{c}{$12P_{3/2}$}\\
    \hline
    DHF                    & 6.6710 & 1.5345[1] & 1.6049 & 6.4748[-1] & 3.7372[-1] & 2.5345[-1] & 1.8800[-1] \\
    RPA(DHF)              & 6.7122 & 1.5227[1] & 1.5340 & 5.9752[-1] & 3.3579[-1] & 2.2328[-1] & 1.6321[-1] \\
    BO                     & 6.4422 & 1.4234[1] & 1.7498 & 7.5067[-1] & 4.4817[-1] & 3.1033[-1] & 2.3333[-1] \\
    RPA(BO)               & 6.4699 & 1.4118[1] & 1.6799 & 7.0160[-1] & 4.1100[-1] & 2.8083[-1] & 2.0914[-1] \\
    SD                     & 6.4235 & 1.4237[1] & 1.6439 & 6.7941[-1] & 3.9582[-1] & 2.6957[-1] & 2.0036[-1] \\
    CCSD                   & 6.4975 & 1.4299[1] & 1.6584 & 6.8588[-1] & 3.9967[-1] & 2.7220[-1] & 2.0228[-1] \\
    CCSDpT                 & 6.4972 & 1.4290[1] & 1.6595 & 6.8662[-1] & 4.0018[-1] & 2.7258[-1] & 2.0258[-1] \\
    CCSDvT                 & 6.4973 & 1.4318[1] & 1.6325 & 6.6860[-1] & 3.8704[-1] & 2.6239[-1] & 1.9435[-1] \\
    CCSDpTvT               & 6.4969 & 1.4309[1] & 1.6336 & 6.6935[-1] & 3.8755[-1] & 2.6277[-1] & 1.9465[-1] \\
    \hline     
    \multicolumn{8}{c}{Other corrections}\\  
    Scaling                &-0.0106	&-0.0018[1] & 0.0043 & 0.0234[-1] & 0.0094[-1] & 0.0139[-1] & 0.0108[-1] \\
    Dressing               & 0.0001 & 0.0001[1] & 0.0005 & 0.0032[-1] & 0.0024[-1] & 0.0019[-1] & 0.0016[-1] \\
    Breit                  & 0.0015 &-0.0001[1] & 0.0010 & 0.0056[-1] & 0.0038[-1] & 0.0028[-1] & 0.0022[-1] \\
    QED                    &-0.0054 & 0.0010[1] &-0.0047 &-0.0284[-1] &-0.0195[-1] &-0.0146[-1] &-0.0115[-1] \\
    Basis extrapolation    &-0.0001 &-0.0004[1] & 0.0008 & 0.0055[-1] & 0.0039[-1] & 0.0029[-1] & 0.0023[-1] \\
    \hline
    Final result           & 6.4824(55) & 1.4297(10)[1] & 1.6355(25) & 6.703(14)[-1] & 3.8755(73)[-1] & 2.6346(81)[-1] & 1.9519(63)[-1] \\
    Uncertainty (\%)       & 0.085 & 0.067 & 0.16 & 0.21 & 0.19 & 0.31 & 0.32\\
    \hline
    Other results           & 6.474(23) \footnote{\label{Roberts2022-4}Roberts \textit{et al.}~(2022),~Ref.~\cite{Roberts2022}} & 1.4303(33) [1]\footnotemark[1] &&&&&\\ 
    					   & 6.480(19)\footnote{\label{Safronova2016-4}Safronova \textit{et al.}~(2016),~Ref.~\cite{Safronova2016}} & 1.4323(61)[1]\footnotemark[2] & 1.620(35)\footnotemark[2] & 6.80(14)[-1]\footnotemark[2] & 3.962(88)[-1]\footnotemark[2] & 2.698(65)[-1]\footnotemark[2] & 2.006(73)[-1]\footnotemark[2]\\
                           & 6.479\footnote{\label{Safronova1999-4}Safronova \textit{et al.}~(1999),~Ref.~\cite{Safronova1999}} & 1.4323[1]\footnotemark[3] &&&&&\\
                           & 6.470\footnote{\label{Blundell1992-4}Blundell \textit{et al.}~(1992),~Ref.~\cite{Blundell1992}} & 1.4293[1]\footnotemark[4] &&&&&\\
                           & 6.507\footnote{\label{Dzuba1989-4}Dzuba \textit{et al.}~(1989),~Ref.~\cite{dzuba1989summation}} & 1.4295[1]\footnotemark[5] &&&&&\\
    \hline
    Experiment             & 6.489(5)\footnote{Toh \textit{et al.}~(2019),~Ref.~\cite{Toh2019Mels}} & 1.4344(7)[1]\footnote{Toh \textit{et al.}~(2019),~Ref.~\cite{Toh2019Polarizabilities}} &&&&&\\
                           & 6.479(31)\footnote{Bouchiat \textit{et al.}~(1984),~Ref.~\cite{Bouchiat1984}} & 1.4320(20)[1]\footnote{Bennett \textit{et al.}~(1999),~Ref.~\cite{PhysRevA.59.R16}} &&&&&\\
    Weighted average       & 6.488(5) & 1.4341(7)[1]                                               &&&&&\\
    \hline
    Difference (\%)         & -0.10 & -0.31 &&&&&\\
    Difference ($\sigma$)   & -0.85  & -3.8 &&&&&\\
    \hline\hline
    \end{tabular}
    \caption{Reduced electric-dipole matrix elements $\brad{7S_{1/2}}D\ketd{n'P_{3/2}}$ (in atomic units a.u.) for $n'=6-12$ in Cs with various approximations. See Table~\ref{tab:Energy_S1} caption for explanation of entries. The notation $x[y]$ stands for $x\times 10^y$.}\label{tab:E1diople_7S1nP3}
\end{table*}

\begin{table*}[ht]
    \centering
    \begin{tabular}{ld{8}d{8}d{6}d{6}d{6}d{6}d{6}}
    \hline\hline
    \multicolumn{1}{c}{}& \multicolumn{1}{c}{6}&\multicolumn{1}{c}{7}&\multicolumn{1}{c}{8}&\multicolumn{1}{c}{9}&\multicolumn{1}{c}{10}& \multicolumn{1}{c}{11}& \multicolumn{1}{c}{12}\\
    \hline
    DHF                    & 0.99499 & 1.3215 & 1.5101 & 1.6344 & 1.7186 & 1.7776 & 1.8198 \\
    RPA(DHF)              & 0.99691 & 1.5070 & 2.3624 & 4.4622 & 16.209 & 14.506 & 5.7621 \\
    BO                     & 0.99146 & 1.2848 & 1.4046 & 1.4706 & 1.5107 & 1.5368 & 1.5549 \\
    RPA(BO)               & 0.99426 & 1.4230 & 1.7814 & 2.1191 & 2.4263 & 2.6977 & 2.9324 \\
    SD                     & 0.99465 & 1.4326 & 1.8087 & 2.1687 & 2.5008 & 2.7963 & 3.0518 \\
    CCSD                   & 0.99455 & 1.4264 & 1.7946 & 2.1443 & 2.4650 & 2.7497 & 2.9971 \\
    CCSDpT                 & 0.99452 & 1.4259 & 1.7927 & 2.1406 & 2.4593 & 2.7419 & 2.9873 \\
    CCSDvT                 & 0.99526 & 1.4781 & 2.0111 & 2.6896 & 3.5480 & 4.6243 & 5.9676 \\
    CCSDpTvT               & 0.99521 & 1.4771 & 2.0079 & 2.6819 & 3.5321 & 4.5939 & 5.9137 \\
    \hline     
    \multicolumn{8}{c}{Other corrections}\\
    Scaling                &-0.00006 &-0.0059 &-0.0242 &-0.0454 &-0.0962 &-0.2396 &-0.3221\\

    Dressing               &-0.00001 &-0.0010 &-0.0050 &-0.0141 &-0.0325 &-0.0628 &-0.1184 \\
    Breit                  & 0.00005 &-0.0087 &-0.0251 &-0.0551 &-0.1066 &-0.1831 &-0.3124 \\
    QED                    & 0.00004 & 0.0055 & 0.0198 & 0.0501 & 0.1058 & 0.1927 & 0.3467 \\
    Basis extrapolation    & 0.00000 &-0.0013 &-0.0051 &-0.0131 &-0.0280 &-0.0519 &-0.0949 \\
    \hline
    Final result           & 0.99523(4) & 1.4656(55) & 1.968(18) & 2.604(38) & 3.375(78) & 4.25(16) & 5.41(25)\\
    Uncertainty (\%)       & 0.0040 & 0.37 & 0.93 & 1.5 & 2.3 & 3.8 & 4.5\\
    \hline
    Other results          & 0.9951(1) \footnote{\label{Roberts2022-5}Roberts \textit{et al.}~(2022),~Ref.~\cite{Roberts2022}} & 1.464(25) [1]\footnotemark[1] &&&&&\\
    					   & 0.995\footnote{\label{Safronova2016-5}Safronova \textit{et al.}~(2016),~Ref.~\cite{Safronova2016}} & 1.425\footnotemark[2] & 1.791\footnotemark[2] & 2.138\footnotemark[2] & 2.452\footnotemark[2] & 2.737\footnotemark[2] & 2.988\footnotemark[2]\\
                           & 0.995\footnote{\label{Safronova1999-5}Safronova \textit{et al.}~(1999),~Ref.~\cite{Safronova1999}} & 1.460\footnotemark[3] & 1.903\footnotemark[3] &&&&\\					
                           & 0.995\footnote{\label{Blundell1992-5}Blundell \textit{et al.}~(1992),~Ref.~\cite{Blundell1992}} & 1.455\footnotemark[4] & 1.940\footnotemark[4] &&&&\\
                           & 0.995\footnote{\label{Dzuba1989-5}Dzuba \textit{et al.}~(1989),~Ref.~\cite{dzuba1989summation}} & 1.499\footnotemark[5] &&&&&\\
    \hline
    Experiment             & 0.99521(23)\footnote{Rafac \& Tanner (1998),~Ref.~\cite{RafTan98}} & 1.4599(28)\footnote{Damitz \textit{et al.}~(2019),~Ref.~\cite{Damitz2019}} & 2.06(15)\footnotemark[19] & 2.52(33)\footnotemark[19] & 3.15(22)\footnotemark[19] & 3.91(39)\footnotemark[19] & 4.49(30)\footnotemark[19] \\
                           & 0.9952(6)\footnote{Amiot \textit{et al.}~(2002),~Ref.~\cite{Amiot2002}} & 1.4654(86)\footnote{Antypas \& Elliot~(2013),~Ref.~\cite{Antypas2013}}&&&&&\\
                           & 0.9952(11)\footnote{Patterson \textit{et al.}~(2015),~Ref.~\cite{Patterson2015}} & 1.502(17)\footnote{Vasilyev \textit{et al.}~(2002),~Ref.~\cite{Vasilyev2002}} &&&&&\\
                           & 0.9941(14)\footnote{Young \textit{et al.}~(1994),~Ref.~\cite{Young1994}} & 1.485(18)\footnote{Borvak (2015),~Ref.~\cite{borvak2015direct}}&&&&&\\
                           & 0.9952(15)\footnote{Derevianko \& Porsev~(2002),~Ref.~\cite{Derevianko2002}} & 1.451(27)\footnote{Shabanova \textit{et al.}~(1979),~Ref.~\cite{ShabMonKhl79}} &&&&&\\
                           & 0.9961(18)\footnote{Rafac \textit{et al.}~(1999),~Ref.~\cite{Rafac1999}} & 1.489(74)\footnotemark[19] &&&&&\\  & 0.9952(23)\footnote{Amini \textit{et al.}~(2003),~Ref.~\cite{Amini2003}} &&&&&&\\
                           & 0.9950(29)\footnote{Zhang \textit{et al.}~(2013),~Ref.~\cite{Zhang2013}} &&&&&&\\
                           & 0.994(1)\footnote{\label{Mor2000}Morton~(2000),~Ref.~\cite{Morton_2000}} &&&&&&\\
                           & 0.995(1)\footnote{Gregoire \textit{et al.}~(2015),~Ref.~\cite{Gregoire2015}} &&&&&&\\
                           & 0.995(14)\footnote{Bouloufa \textit{et al.}~(2007),~Ref.~\cite{Bouloufa2007}} &&&&&&\\
    Weighted average       & 0.99517(20) & 1.4619(26) &&&&&\\
    \hline
    Difference (\%)         & 0.006 & 0.25 &-4.6  & 3.4  & 7.1 & 8.7  & 21 \\
    Difference ($\sigma$)   & 0.28 & 0.61 &-0.64 & 0.26 & 0.98 & 0.81 & 2.7 \\
    \hline\hline
    \end{tabular}
    \caption{Normalized ratios of electric-dipole matrix elements $\xi_{6n'}\equiv(1/\sqrt{2})\brad{6S_{1/2}}D\ketd{n'P_{3/2}}/\brad{6S_{1/2}}D\ketd{n'P_{1/2}}$ for $n'=6-12$ in Cs in various approximations.
See Table~\ref{tab:Energy_S1} caption for explanation of entries. In the nonrelativistic limit, $\xi_{nn'}=1$ whereas relativistic and many-body effects cause $\xi_{nn'}$ to deviate significantly from 1.}\label{tab:ratio_6}
\end{table*}

\begin{table*}[ht]
    \centering
    \begin{tabular}{ld{9}d{7}d{6}d{6}d{6}d{6}d{6}}
    \hline\hline
    \multicolumn{1}{c}{}& \multicolumn{1}{c}{6}&\multicolumn{1}{c}{7}&\multicolumn{1}{c}{8}&\multicolumn{1}{c}{9}&\multicolumn{1}{c}{10}& \multicolumn{1}{c}{11}& \multicolumn{1}{c}{12}\\
    \hline
    DHF                    & 1.0689 & 0.98561 & 1.2319 & 1.3577 & 1.4415 & 1.5003 & 1.5429 \\
    RPA(DHF)              & 1.0667 & 0.98593 & 1.2481 & 1.4048 & 1.5244 & 1.6180 & 1.6916 \\
    BO                     & 1.0860 & 0.98069 & 1.2274 & 1.3313 & 1.3926 & 1.4324 & 1.4597 \\
    RPA(BO)               & 1.0833 & 0.98112 & 1.2423 & 1.3682 & 1.4506 & 1.5083 & 1.5501 \\
    SD                     & 1.0827 & 0.98188 & 1.2512 & 1.3861 & 1.4765 & 1.5405 & 1.5872 \\
    CCSD                   & 1.0810 & 0.98188 & 1.2466 & 1.3788 & 1.4671 & 1.5298 & 1.5756 \\
    CCSDpT                 & 1.0811 & 0.98179 & 1.2463 & 1.3782 & 1.4662 & 1.5287 & 1.5743 \\
    CCSDvT                 & 1.0803 & 0.98218 & 1.2531 & 1.3944 & 1.4921 & 1.5631 & 1.6161 \\
    CCSDpTvT               & 1.0804 & 0.98214 & 1.2527 & 1.3937 & 1.4910 & 1.5618 & 1.6146 \\
    \hline
    \multicolumn{8}{c}{Other corrections}\\  
    Scaling                & 0.0003 &-0.00019 &-0.0019 &-0.0013 &-0.0027 &-0.0035 & 0.0000 \\
    Dressing               & 0.0000 & 0.00007 &-0.0001 &-0.0003 &-0.0005 &-0.0007 &-0.0008 \\
    Breit                  &-0.0010 & 0.00022 &-0.0038 &-0.0068 &-0.0090 &-0.1144 &-0.0123 \\
    QED                    & 0.0002 & 0.00002 & 0.0021 & 0.0040 & 0.0058 & 0.0057 & 0.0084 \\
    Basis extrapolation    & 0.0000 & 0.00001 &-0.0005 &-0.0009 &-0.0013 &-0.0013 &-0.0020 \\
    \hline
    Final result           & 1.0799(5) & 0.9823(1) & 1.2485(22) & 1.3885(36) & 1.4833(50) & 1.5523(60) & 1.6080(66) \\
    Uncertainty            & 0.049 & 0.014 & 0.18 & 0.26 & 0.33 & 0.39 & 0.41 \\
    \hline                 
    Other results          & 1.0800(11) \footnote{\label{Roberts2022-6}Roberts \textit{et al.}~(2022),~Ref.~\cite{Roberts2022}} & 0.9822(1)\footnotemark[1] &&&&&\\
    					   & 1.0797\footnote{\label{Safronova2016-6}Safronova \textit{et al.}~(2016),~Ref.~\cite{Safronova2016}} & 0.9823\footnotemark[2] & 1.2531\footnotemark[2] & 1.3795\footnotemark[2] & 1.4683\footnotemark[2] & 1.5299\footnotemark[2] & 1.5761\\
                           & 1.080\footnote{\label{Safronova1999-6}Safronova \textit{et al.}~(1999),~Ref.~\cite{Safronova1999}} & 0.982\footnotemark[3] &&&&&\\					
                           & 1.080\footnote{\label{Blundell1992-6}Blundell \textit{et al.}~(1992),~Ref.~\cite{Blundell1992}} & 0.982\footnotemark[4] &&&&&\\
                           & 1.082\footnote{\label{Dzuba1989-6}Dzuba \textit{et al.}~(1989),~Ref.~\cite{dzuba1989summation}} & 0.983\footnotemark[5] &&&&&\\
    \hline
    Experiment             & 1.0799(12)\footnote{Toh \textit{et al.}~(2019),~Ref.~\cite{Toh2019Mels}} &&&&&&\\
    \hline\hline
    \end{tabular}
    \caption{Normalized ratios of electric-dipole matrix elements $\xi_{7,n'}\equiv(1/\sqrt{2})\brad{7S_{1/2}}D\ketd{n'P_{3/2}}/\brad{7S_{1/2}}D\ketd{n'P_{1/2}}$ for $n'=6-12$ in Cs in various approximations. See Table~\ref{tab:Energy_S1} caption for explanation of entries. In the nonrelativistic limit, $\xi_{nn'}=1$ whereas relativistic and many-body effects cause $\xi_{nn'}$ to deviate significantly from 1.}\label{tab:ratio_7}
\end{table*}

\section*{Acknowledgements}
We thank B. M. Roberts,  C. J. Fairhall, and J. S. M. Ginges for providing QED corrections and useful discussions. 
This work was supported in part by the U.S. National Science Foundation Grants No.~PHY-1912465 and No.~PHY-2207546, by the Sara Louise Hartman Endowed Professorship in Physics, and by the Center for Fundamental Physics at Northwestern University. 

\appendix
\newpage 
\section{Details of constructing the \texorpdfstring{$B$-spline}{} finite basis set}
\label{App:Finite-Basis-Set}
The $B$-spline basis set is one example of the finite basis sets, the workhorse of numerous atomic structure and quantum chemistry codes. The $B$-spline basis set was popularized by the Notre Dame group~\cite{Johnson1988,SapJoh1996-BsplineReview,WRJBook} and since then has found numerous applications in high-precision relativistic atomic-structure calculations, especially those based on many-body perturbation theory (MBPT). The power of the finite basis sets lies in the ability to carry out summations over intermediate single-particle orbitals. Such summations are ubiquitous in numerical implementations of MBPT formalism. Since an exact atomic single-particle spectrum consists of a numerable yet infinite set of bound states and an innumerable set of states in the continuum, the combined set contains an infinite number of eigenfunctions and is simply impractical in numerical implementations. 
A finite basis set is a numerical approximation to the exact eigenspectrum, replacing it with a numerically complete yet finite-sized set. 

The procedure for constructing a finite basis set is as follows. First, we confine the atom to a spherical cavity of radius $R_{\mathrm{max}}$. Then the exact single-particle spectrum becomes countable as the continuum is discretized, yet the confined atomic spectrum still contains an infinite number of eigenstates. To make the basis finite, the orbitals are expanded over a numerically complete set of support polynomials, the B-splines in our case~\cite{Johnson1988,SapJoh1996-BsplineReview,WRJBook}. Finally, the single-particle Dirac Hamiltonian is diagonalized in this finite-sized Hilbert space, producing the desired eigenspectrum that is now finite.  

One of the technical drawbacks of the original Notre Dame $B$-spline implementation~\cite{Johnson1988} is the occurrence of the so-called spurious states, which do not map into the physical states of the Hamiltonian. This drawback was rectified with the introduction of the dual-kinetic balance (DKB) boundary conditions~\cite{Shabaev2004} for $B$-spline sets. The original work~\cite{Shabaev2004} focused on the hydrogenlike systems and then the DKB construction was extended to DHF potentials for multielectron atoms~\cite{BELOY2008}. In our calculations, we use the DKB $B$-spline basis sets described in Ref.~\cite{BELOY2008}.     

In this paper, we carry out computations for uncharacteristically large principle quantum numbers (up to $n=12$). In the basis-set construction described above, even if the cavity radius is large, $R_{\mathrm{max}} \gg a_0$, only the lower energy orbitals map into the single-particle states of an unconfined atom, with higher-energy orbitals no longer fitting into the cavity. Then the mapping of basis-set orbitals to ``physical'' orbitals corresponding to an unconfined atom becomes spoiled. We now discuss our strategy for selecting $R_{\mathrm{max}}$.

To ensure the correct mapping of the basis-set orbitals to physical ones, we carry out a supporting finite-difference calculation. The starting point of our calculation is the frozen-core DHF method. The finite-difference method is based on a numerical integration of the DHF equation on a sufficiently large grid to fully accommodate the desired atomic DHF orbitals~\cite{WRJBook}. In other words, the finite-difference method provides the reference results for an unconfined atom. The $B$-spline basis-set construction method solves the same DHF problem but in a cavity. We vary the cavity radius $R_{\mathrm{max}}$ and compare the energies and other supporting quantities, such as the transition amplitudes, of our target basis-set orbitals with the finite-difference results.  

\begin{figure}[ht!]
      \includegraphics[width=\columnwidth]{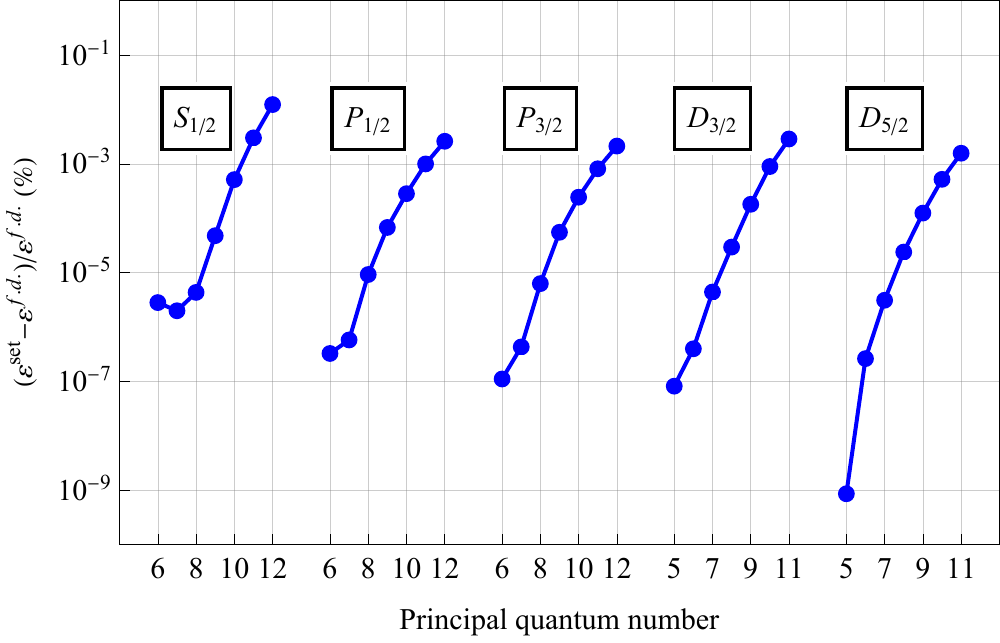}
      \caption{Fractional differences $\left| (\varepsilon^\mathrm{set} - \varepsilon^\mathrm{f.d.})/\varepsilon^\mathrm{f.d.} \right|$  between the DKB $B$-spline basis set and finite-difference (f.d.) Dirac-Hartree-Fock eigenenergies $\varepsilon_i$ for several angular symmetries as functions of the principle quantum number. Basis-set parameters: number of splines for a fixed angular symmetry is  $M=60$, cavity radius $R_{\max}=250\,\mathrm{a.u.}$ and $B$-spline order $k=9$. }
      \label{fig:per_of_diff}
  \end{figure}

We presented such a comparison for the DHF eigenenergies in Fig.~\ref{fig:per_of_diff}. Here the $B$-spline basis set contains $M=60$ basis functions per partial wave ($B$-splines of order $k=9$) generated in a cavity of radius $R_{\mathrm{max}}=250\,\mathrm{a.u.}$ We plot the fractional difference between the
basis-set and finite-difference eigenenergies and find the agreement to be better than $0.015\%$. As expected, the difference between the basis set and ``physical'' orbital energies worsens with increasing principle quantum numbers, i.e., with the increasing spatial extent of the orbitals.
By using the same basis set, we also investigated the electric- and magnetic-dipole matrix elements between atomic orbitals. The basis-set values for the matrix elements differ from their finite-difference counterparts by up to $0.1\%$ for orbitals involving large $n$ values. We notice that a larger cavity radius, such as $R_{\mathrm{max}}=500\, \mathrm{a.u.}$, does not necessarily lead to a better numerical accuracy, because the resulting larger grid step size results in a poorer $B$-spline grid coverage. To fix this, one should increase $M$, the number of B-splines used in the basis-set generation, as well as the $B$-spline order $k$. Increasing $M$, however, leads to larger basis sets and, thereby, a polynomial $M^\gamma$ increase in computational time (the power $\gamma$ depends on a specific many-body scheme, with the steepest scaling for our most sophisticated CC calculations). The same observation applies for $k$. Our basis set, as described in Sec.~\ref{Sec:Numericals} of the main text, is a compromise between numerical accuracy and computational time. To further improve our numerical accuracy in basis-set-based many-body calculations, we replace the lowest-order DHF values of matrix elements with those computed using finite-difference orbitals.

\section{Constructing finite basis set of  Brueckner orbitals}\label{App:BO}

An introduction to the Brueckner orbital (BO) method was given in Sec.~\ref{Sec:BO} of the main text. The second-order expression for the self-energy operator $\Sigma$ was given by Eq.~\eqref{Eq:SE-II}.
The goal of this Appendix is to show that one may generate a BO basis set from a DHF finite basis set, described in Appendix~\ref{App:Finite-Basis-Set}, by using basis rotation. Such a generated BO set retains all the numerically useful properties of the original DHF basis set, but has the extra advantage of producing important third-order correlation corrections unobtainable with a DHF basis set in an RPA calculation, as discussed in Sec.~\ref{Sec:RPA}. 

By construction, the DHF basis orbitals $v_k$ satisfy the eigenvalue equation 
\begin{equation}
h_0 v_{k}=\varepsilon_{k}^\mr{DHF} v_{k} \,.
\end{equation}
The DHF basis is numerically compete, orthonormal, and of finite size $M$. Notice that we attached the label DHF to the energies; in Eq.~(\ref{Eq:SE-II}) that label was suppressed. We would now like to find solutions to the BO eigenvalue equation with the self-energy operator $\Sigma$ included
\begin{equation}
\left(  h_0 +\Sigma\right) u=\varepsilon u \,. \label{Eq:BO-eigval-eqn}
\end{equation}
Since the DHF set $\{v_{k} \}$ is numerically complete, we can expand the solution $u$ in terms of the DHF basis orbitals as $u =\sum_{k}c_{k}v_{k}$. 
By plugging this expansion into Eq.~(\ref{Eq:BO-eigval-eqn}) and using the orthonormality  of $\{v_{k} \}$, we arrive at 
\begin{equation}
\sum_{k}\left(  \varepsilon_{k}^\mr{DHF}\delta_{mk}+
\Sigma_{mk}\right)c_{k} =\varepsilon^\mr{BO}\sum_{k}c_{k}\delta_{mk}\,,
\end{equation}
which may be cast in matrix form as 
\begin{equation}\label{Eq:BO-eigenValEq} 
M^{\rm BO}\mb{c}=\varepsilon^\mr{BO}\mb{c} \,,
\end{equation}
where
\begin{equation}
    M^{\rm BO}=
\begin{pmatrix}
\Sigma_{11}+\varepsilon_{1}^\mr{DHF} & \Sigma_{12}                          & \cdots & \Sigma_{1M} \\
\Sigma_{21}                          & \Sigma_{22}+\varepsilon_{2}^\mr{DHF} & \cdots & \Sigma_{2M}\\
\vdots                               & \vdots                               & \ddots & \vdots \\
\Sigma_{M1}                          & \Sigma_{M2}                          & \cdots & \Sigma_{MM}+\varepsilon_{M}^\mr{DHF}%
\end{pmatrix}\,.
\end{equation}
By solving this equation we find the $M$ BO eigenvalues $\varepsilon^\mr{BO}$ and the corresponding eigenvectors of expansion coefficients $\bf c$. Using these expansion coefficients we can assemble the desired Brueckner orbitals as $u =\sum_{k}c_{k}v_{k}$.

The numerical implementation of this method may be significantly sped up by using angular reduction as follows. We begin by writing
\begin{subequations}\label{eq:Ang_Dec}
    \begin{align}
        g_{ijkl}&=\sum_LJ_L(ijkl)X_L(ijkl)\,,\\
        \tilde{g}_{ijkl}&=\sum_LJ_L(ijkl)Z_L(ijkl)\,.
    \end{align}
\end{subequations}
where
\begin{align}
    J_L(ijkl)&\equiv\sum_{M}(-1)^{j_i-m_i+j_j-m_j}\nonumber\\
    &\times\begin{pmatrix}
        j_i & L & j_k \\
        -m_i & -M & m_k
        \end{pmatrix}\begin{pmatrix}
        j_j & L & j_l \\
        -m_j & M & m_l
        \end{pmatrix}\,.
\end{align}

The quantity $X_L(ijkl)$ is expressed in terms of the reduced matrix element of the normalized spherical harmonic  $C_L(\hat{\bf r})$ and the Slater integral $R_L(ijkl)$ as
\begin{equation}\label{PP_redux_X}
     X_L(ijkl) = (-1)^L\rmel{\kappa_i}{C_L}{\kappa_k}\rmel{\kappa_j}{C_L}{\kappa_l}R_L(ijkl)\,,
\end{equation}
where $\kappa$ is the relativistic angular quantum number that uniquely encodes both the total angular momentum $j$ and the orbital angular momentum $\ell$, $\kappa=(\ell-j)\left(  2j+1\right)$. The quantity $Z_L(ijkl)$ may be expressed in terms of $X_L(ijkl)$ via the recoupling formula
\begin{align}\label{PP_redux_Z}
    Z_L(ijkl) &= X_L(ijkl) \nonumber\\
              &+ \sum_{L'}[L]\begin{pmatrix}
                j_k & j_i & L \\
                j_l & j_j & L'
    \end{pmatrix}X_{L'}(ijlk)\,,
\end{align}
where $[L]\equiv2L+1$ and $\begin{pmatrix}j_k & j_i & L \\j_l & j_j & L'\end{pmatrix}$ is the $6j$-symbol. 

Using the angular decompositions~\eqref{eq:Ang_Dec}, we may write the matrix elements of the second-order self-energy operator~(\ref{Eq:SE-II}) as
\begin{align}\label{Eq:SE-II-angular} 
\Sigma_{ij}&=\delta_{\kappa_{i}\kappa_{j}}\delta_{m_{i}m_{j}} \nonumber\\
&\times\left( \sum_{amn, L} 
\frac{(-1)^{j_m+j_n+j_a+j_i}}{[L,j_i]}
\frac{X_{L}(aimn)  Z_{L}(mnaj)}
{\varepsilon_{a0}-\varepsilon_{mn}} \right. \nonumber\\ 
 &+  \left.\sum_{abm}\frac{\left(  -1\right)
 ^{j_m+j_i+j_a+j_b}}{\left[L,j_i\right]  }\frac{Z_{L}\left(  miab\right)
 X_{L}\left(  abmj\right)  }{\varepsilon_{m0}-\varepsilon_{ab}} \right)\,.  
\end{align}
where $[L,j_i] \equiv [L][j_i] = (2L+1) (2j_i+1)$. Notice the angular selection rules enforced by the $\delta$ symbols, reflecting the fact that the self-energy operator is a scalar. The matrix $M^{\rm BO}$ may then be rearranged into a block-diagonal form, with each block corresponding to a different $\kappa$ value. Solving the eigenvalue equation~\eqref{Eq:BO-eigenValEq} is thus equivalent to diagonalizing these blocks individually.

To speed up the computations of the self-energy matrix elements~\eqref{Eq:SE-II-angular} further, we introduce a kernel (here and below the angular symmetry of the block is fixed)
\begin{equation}
K(r,r^{\prime})=\left(
\begin{array}
[c]{cc}%
K_{PP}\left(  r,r^{\prime}\right)  & K_{PQ}\left(  r,r^{\prime}\right) \\
K_{QP}\left(  r,r^{\prime}\right)  & K_{QQ}\left(  r,r^{\prime}\right)
\end{array}
\right) \, ,
\end{equation}
so that the self-energy matrix elements can be assembled from the large ($P$) and small ($Q$) components of the Dirac bispinors as
\begin{equation}
\Sigma_{ij}=\int dr dr^{\prime}
\left(  P_{i}\left(  r\right), Q_{i}\left(
r\right)  \right)  K(r,r^{\prime})\binom{P_{j}\left(  r^{\prime}\right)
}{Q_{j}\left(  r^{\prime}\right)} \,. \label{Eq:BO-melsOfSEwithKernel}
\end{equation}

A straightforward but somewhat tedious derivation results in the following expressions for the kernels 
($X$ and $Y$ stand either for $P$ or $Q$)
\begin{align}
K_{XY}\left(  r,r^{\prime}\right)  &= 
\frac{1}{\left[  L,j_i\right]} \left(
\sum_{amn,L}
\frac{K_{XY}^{\left(  amn, L\right)}\left(  r,r^{\prime}\right)  }%
{\varepsilon_{a0}-\varepsilon_{mn}}\right. \nonumber \\
& +\left.\sum_{abm,L}\frac{K_{XY}^{\left(  abm, L\right)}\left(  r,r^{\prime}\right)  }%
{\varepsilon_{m0}-\varepsilon_{ab}} \right) \,. \label{Eq:BO-Kernel}
\end{align}
Remember that the angular symmetry here is fixed, $\kappa_i=\kappa_j$. The two subkernels appearing in the sums are, explicitly,
\begin{align}\label{Eq:BO-sub-Kernel-amn} 
K_{XY}^{\left(  amn, L \right)  }\left(  r,r^{\prime}\right) 
&=H_{L}\left(mnai\right)  X_{n}\left(  r\right)          v_{L}\left(  am;r\right)  \nonumber\\ 
&\times\left\{ H_{L}(mnaj)Y_{n}\left(  r^{\prime}\right) v_{L}\left(  am;r^{\prime}\right)  \right.\nonumber\\
& +[L]\sum_{L^{\prime}}H_{L^{\prime}}(nmaj)\left\{
\begin{array}
[c]{ccc}%
j_n & j_j & L\\
j_m & j_a & L^{\prime}%
\end{array}\right\}\nonumber\\
&\times\left.  Y_{m}\left(  r^{\prime}\right)  v_{L^{\prime}}\left(  an;r^{\prime
}\right)  \right\}\,,
\end{align}%
and 
\begin{align}\label{Eq:BO-sub-Kernel-abm} \
K_{XY}^{\left(  abm,L\right)  }\left(  r,r^{\prime}\right)   
&=H_{L}\left(  mjab\right)Y_{b}\left(r^{\prime}\right) v_{L}\left(  am;r^{\prime}\right)  \nonumber \\
&\times\left\{
H_{L}\left(  miab\right)  X_{b}\left(  r\right)  v_{L}\left(  am;r\right)\right.\nonumber\\ 
&+[L]\sum_{L^{\prime}}H_{L^{\prime}}(imab)\left\{
\begin{array}
[c]{ccc}%
j_i & j_b & L\\
j_m & j_a & L^{\prime}%
\end{array}
\right\} \nonumber\\
&\times\left.X_{a}\left(  r\right)  v_{L^{\prime}}\left(  bm;r\right)  
\right\} \,,
\end{align}
where
\begin{equation}
H_{L}(abcd)=(-1)^{L}\langle\kappa_{a}||C_L||\kappa_{c}\rangle\langle\kappa_{b}||C_L||\kappa_{d}\rangle\,,
\end{equation}
and $v_{L}\left(  bm;r\right)$ is the screening potential,
\begin{equation*}
    v_{k}\left(  ij;r \right) = \int dr' \frac{r^k_<}{r^{k+1}_> } \left[ P_i(r') P_j(r') + Q_i(r') Q_j(r') \right] \,,
\end{equation*}
with the conventional definitions $r_< = \min(r,r')$ and $r_> = \max(r,r')$.

To reiterate, we start by generating the DHF finite basis set on a certain radial grid, as described in the main text. Then, for a fixed symmetry $\kappa$, we tabulate the four elements of the kernel~(\ref{Eq:BO-Kernel}) on the grid with the help of 
formulas~\eqref{Eq:BO-sub-Kernel-amn} and~\eqref{Eq:BO-sub-Kernel-abm}). All the evaluations are 
carried out with the DHF finite basis set. This is the most time-consuming part of the calculations. Then, with the tabulated kernel, we compute the matrix elements~(\ref{Eq:BO-melsOfSEwithKernel}) of the self-energy matrix and solve the eigenvalue equation~(\ref{Eq:BO-eigenValEq}). This provides us with the desired BO spectrum and corresponding eigenvectors of expansion coefficients of the BO orbitals over the original DHF basis. We normalize these eigenvectors to guarantee that the BO basis set is orthonormal. Finally, with these expansion coefficients, we assemble the BO basis-set functions.   

Finally, we turn to the question of how to choose the reference energy $\varepsilon_0$. Since we are diagonalizing the BO Hamiltonian $H_\mr{DHF} + \Sigma$ for each individual angular symmetry $\kappa$ ($s_{1/2}, p_{1/2}, p_{3/2}, \ldots $), we can pick   
different values of $\varepsilon_0$ for different $\kappa$. However, within each $\kappa$ block, $\varepsilon_0$ is fixed. Because we are interested in the low-energy valence states, in the calculations reported in this paper, we fix $\varepsilon_0$ to the lowest valence electron DHF energy for a given $\kappa$, e.g., for Cs $s_{1/2}$ states we pick $\varepsilon_0=\varepsilon_{6s}$ and for $p_{1/2}$ states we pick 
$\varepsilon_0=\varepsilon_{6p_{1/2}}$, and so on.

\section{Finite-basis-set implementation of the random phase approximation}
\label{App:RPA}
An introduction to the random phase approximation (RPA) can be found in Sec.~\ref{Sec:RPA}. The focus of this appendix is to describe an efficient numerical finite-basis-set implementation of the RPA method. 
As a starting point, we reproduce formula~\eqref{RPA_core} from the main text. We are interested in computing matrix elements of a one-electron operator $Z=\sum_k z_k$,
where the sum goes over all the electrons. The RPA-dressed matrix elements (vertices) are
\begin{subequations}\label{Eq:ZRPA}
    \begin{align}
Z_{ma}^\mathrm{RPA}  &  =z_{ma}\nonumber\\
&+\sum_{bn}\left(  \frac{Z_{bn}^\mathrm{RPA}\tilde{g}%
_{mnab}}{\varepsilon_{b}-\varepsilon_{n}-\omega}+\frac{Z_{nb}^\mathrm{RPA}%
\tilde{g}_{mban}}{\varepsilon_{b}-\varepsilon_{n}+\omega}\right)
\,,\\
Z_{am}^\mathrm{RPA}  &  =z_{am}\nonumber\\
&+\sum_{bn}\left(  \frac{Z_{bn}^\mathrm{RPA}\tilde{g}%
_{anmb}}{\varepsilon_{b}-\varepsilon_{n}-\omega}+\frac{Z_{nb}^\mathrm{RPA}%
\tilde{g}_{abmn}}{\varepsilon_{b}-\varepsilon_{n}+\omega}\right)\,,
\end{align}
\end{subequations}
where $\omega$ is the frequency of the perturbation driving the transition, which, in our case, the $w \rightarrow v$ transition: $\omega \equiv  \varepsilon_w - \varepsilon_v$. Notice that these RPA-dressed matrix elements are defined between the core ($a$) and the excited ($m$) orbitals. The matrix elements between the two valence orbitals are given by the second-order expression in terms of the above RPA-dressed matrix elements, 
\begin{align}\label{RPA_valence-1}
		Z_{wv}^\mathrm{RPA} &= z_{wv} \nonumber\\
  &+ \sum_{an}\frac{Z^{\rm RPA}_{am}\tilde{g}_{wmva}}{\varepsilon_a-\varepsilon_m-\omega}+ \sum_{am}\frac{\tilde{g}_{wvma}Z^{\rm RPA}_{ma}}{\varepsilon_a-\varepsilon_m+\omega}\,.
\end{align}

Clearly, we need to first find the RPA-dressed vertices $Z_{ma}^\mathrm{RPA}$ and $Z_{am}^\mathrm{RPA}$. Usually, the set of equations~(\ref{Eq:ZRPA}) is solved iteratively (see e.g., Ref.~\cite{Johnson1996}), with subsequent iterations recovering higher and higher orders of MBPT. In practical applications, however, sometimes the convergence is poor. Here we present a method to determine the RPA-dressed vertices in one shot, avoiding the iterations altogether. Our method also offers computational advantages when calculating matrix elements for multiple transitions.

We start by defining the following auxiliary quantities
\begin{subequations}
    \begin{align}
\chi_{ma}  &  \equiv\frac{Z_{ma}^\mathrm{RPA}}{\varepsilon_{a}-\varepsilon_{m}%
+\omega}\,,\\
\eta_{ma}^*  &  \equiv\frac{Z_{am}^\mathrm{RPA}}{\varepsilon_{a}-\varepsilon
_{m}-\omega}\,,
\end{align}
\end{subequations}
with which Eqs.~(\ref{Eq:ZRPA}) for the RPA-dressed vertices can be recast into the form
\begin{subequations}\label{Eq:RPAinhom}
    \begin{align}
z_{ma}&=-\left( \varepsilon_{m}-\varepsilon_{a}-\omega\right)  \chi_{ma}\nonumber\\
&-\sum_{bn}\left(  \chi_{nb}\tilde{g}_{bmna}+\eta_{nb}^*\tilde{g}_{nmba}\right)   \,,\\
z_{am}&=-\left(  \varepsilon_{m}-\varepsilon_{a}+\omega\right)  \eta_{ma}^*\nonumber\\
&+\sum_{bn}\left(  \eta_{nb}^*\tilde{g}_{nabm}+\chi_{nb}\tilde{g}_{banm}\right)\,.
\end{align}
\end{subequations}

This system of equations for  $\chi_{ma}$ and $\eta_{ma}^*$ is linear. It is inhomogeneous with the driving term $\left(  -z_{ma},-z_{am}\right)$. We can find the solution of this inhomogeneous set of equations by first solving the eigenvalue problem
\begin{subequations}\label{Eq:RPA-eigenvalEq}
    \begin{align}
\omega\chi_{ma}&=(\varepsilon_{m}-\varepsilon_{a})\chi_{ma}\nonumber\\
&+\sum_{nb}\left(  \ \tilde{g}_{bmna}\chi_{nb}+\ \tilde{g}_{nmba}\eta_{nb}^*\right)\,,\\
\omega\eta_{ma}^*&=(\varepsilon_{m}-\varepsilon_{a})\eta_{ma}^*\nonumber\\
&+\sum_{nb}\left(\tilde{g}_{nabm}\eta_{nb}^*+\ \tilde{g}_{banm}\chi_{nb}\right)\,,
\end{align}
\end{subequations}
to obtain the eigenpair $\left\{  \omega_{\mu},\chi
_{ma}^{\mu},\left(  \eta_{ma}^{\mu}\right)  ^*\right\}$. The eigenfrequencies $\omega_{\mu}$ can be interpreted as frequencies of particle-hole excitations of the atomic closed-shell core. 

There are two relevant properties of the eigensystem~\eqref{Eq:RPA-eigenvalEq}: symmetry and orthonormality. First, by examining Eqs.~(\ref{Eq:RPA-eigenvalEq}), one concludes that for every eigenfrequency $\omega_{\mu}$ there is an eigenfrequency of opposite sign $-\omega_{\mu}$. Second, the two corresponding eigenvectors 
are related: if the triple
$\left\{  \omega_{\mu},\left(  \chi_{ma}^{\mu
},\left(  \eta_{ma}^{\mu}\right)  ^*\right)  \right\}  $ belongs to the eigensystem, so does its  
negative-frequency 
counterpart $\left\{  -\omega_{\mu},\left(  \left(  \eta_{ma}^{\mu}\right)
^*,\chi_{ma}^{\mu}\right)  \right\}$. Further, the eigenvectors satisfy the orthonormality condition
\begin{equation}
\sum_{ma}\left[\chi_{ma}^{\lambda}\left(  \chi_{ma}^{\mu}\right)  ^*-\left(
\eta_{ma}^{\lambda}\right)  ^*\eta_{ma}^{\mu}\right]=\mathrm{sign}\left(
\omega_{\mu}\right)  \delta_{\lambda\mu} \,.
\label{Eq:RPA-orthonormality}
\end{equation}



Once the eigenvalue problem, Eqs.~(\ref{Eq:RPA-eigenvalEq}), is solved and we
obtain a set of eigenvalues $\omega^{\mu}$ and eigenvectors $\left(  \chi
_{ma}^{\mu},\left(  \eta_{ma}^{\mu}\right)  ^*\right)  $, we search for a
solution of the inhomogeneous equations as an expansion over the complete set of
eigenvectors
\begin{equation}
\left(
\begin{array}
[c]{c}%
\chi_{ma}\\
\eta_{ma}^*%
\end{array}
\right)  =\sum_{\mu}c_{\mu}\left(
\begin{array}
[c]{c}%
\chi_{ma}^{\mu}\\
\left(  \eta_{ma}^{\mu}\right)  ^*%
\end{array}
\right) \,. \label{Eq:RPA:expansion}
\end{equation}
Substituting this expansion into Eqs.~(\ref{Eq:RPAinhom}), one
obtains
\begin{equation}
\sum_{\mu}\left(  \omega-\omega_{\mu}\right)  c_{\mu}\left(
\begin{array}
[c]{c}%
\chi_{ma}^{\mu}\\
-\left(  \eta_{ma}^{\mu}\right)  ^*%
\end{array}
\right)  =\left(
\begin{array}
[c]{c}%
z_{ma}\\
z_{am}%
\end{array}
\right)  .
\end{equation}
Multiplying from the right by $\left(  \left(  \chi_{ma}^{\nu}\right)  ^{\ast
},\eta_{ma}^{\nu}\right)$ and using the orthogonality relation~(\ref{Eq:RPA-orthonormality}),
one finds the expansion coefficients
\begin{equation}
c_{\mu}=\frac{\mathrm{sign}\left(  \omega^{\mu}\right)  }{\omega-\omega_{\mu}%
}\sum_{ma}\left[\left(\chi_{ma}^{\mu}\right)  ^*z_{ma}+\eta_{ma}%
^{\mu}z_{am}\right]  .
\end{equation}

Finally, returning to the definitions of $\chi_{ma}$ and $\eta_{ma}$ and introducing
\begin{equation}
S_{\mu}=\sum_{ma}\left[\left(  \chi_{ma}^{\mu}\right)  ^*z_{ma}%
+\eta_{ma}^{\mu}z_{am}\right] \,, \label{Eq:RPA:Smu}
\end{equation}
we arrive at the desired RPA-dressed vertices 
\begin{align}
Z_{ma}^\mr{RPA}  &  =\left(  \varepsilon_{a}-\varepsilon_{m}+\omega\right)
\sum_{\mu}\frac{\mathrm{sign}\left(  \omega^{\mu}\right)}%
{\omega-\omega_{\mu}}S_{\mu}\chi_{ma}^{\mu}\,,\\
Z_{am}^\mr{RPA}  &  =\left(  \varepsilon_{a}-\varepsilon_{m}-\omega\right)
\sum_{\mu}\frac{\mathrm{sign}\left(  \omega^{\mu}\right)}%
{\omega-\omega_{\mu}}S_{\mu}\left(  \eta_{ma}^{\mu}\right)  ^*.
\end{align}

The final step is the angular reduction of the above expressions.
Without losing generality, we assume that the one-electron operator $Z$ is an irreducible tensor operator of rank $J$. We also fix its $M$, the spherical tensor component. We remind the reader that RPA describes particle-hole excitations 
of a closed-shell core. Such excitations by an operator $Z$ necessarily required the excitation total angular momentum $J$ and its projection $M$. The parity of the particle-hole excitation must be the same as of the operator $Z$.
Thereby, we can introduce the following parametrization using the conventional Clebsch-Gordan coefficients, 
\begin{subequations}
    \begin{align}
X_{m\left(  a\rightarrow\kappa_{m}\right)  }  &  =\sum_{m_{m}m_{a}}\left(
-1\right)  ^{j_{a}-m_{a}}C_{j_{m}m_{m}j_{a}-m_{a}}^{JM}\chi_{ma}\,, \\
Y_{m\left(  a\rightarrow\kappa_{m}\right)  }  &  =\sum_{m_{m}m_{a}}\left(
-1\right)  ^{j_{a}-m_{a}+J-M}C_{j_{m}m_{m}j_{a}-m_{a}%
}^{J-M}\eta_{ma}^*\,,
\end{align}
\end{subequations}
where $\left(  a\rightarrow\kappa_{m}\right)$ denotes an excitation
channel, e.g., $1s_{1/2} \rightarrow p_{3/2}$ for the electric-dipole operator ($J=1$ and odd parity). Notice the additional phase factors and negative magnetic quantum numbers for core orbitals, see Ref.~\cite{WRJBook} for justification. The reduced coefficients $X_{m\left(  a\rightarrow\kappa_{m}\right)  }$ and $Y_{m\left(  a\rightarrow\kappa_{m}\right)  }$ no longer depend on the magnetic quantum numbers.

Carrying out the summation over the magnetic quantum
numbers in the eigenvalue equations~(\ref{Eq:RPA-eigenvalEq}), we arrive at their reduced form 
for the coefficients $X_{m\left(  a\rightarrow\kappa_{m}\right)  }$ and $Y_{m\left(  a\rightarrow\kappa_{m}\right)  }$,
\begin{subequations}
    \begin{align}
\omega X_{m\left(  a\rightarrow\kappa_{m}\right)  }&=(\varepsilon_{m}-\varepsilon_{a})X_{m\left(  a\rightarrow\kappa_{m}\right)}\nonumber\\
  &+\sum_{nb}\frac{\left(  -1\right)  ^{J+j_n-j_b}}{\left[  J\right]  }%
Z_{J}(bmna)X_{n\left(  b\rightarrow\kappa_{n}\right)  }\nonumber\\
&  +\sum_{nb}\frac{1}{\left[  J\right]  }Z_{J}(nmba)Y_{n\left(  b\rightarrow
\kappa_{n}\right)  }\,,\\
-\omega Y_{m\left(  a\rightarrow\kappa_{m}\right)}&=(\varepsilon_{m}-\varepsilon_{a})Y_{m\left(  a\rightarrow\kappa_{m}\right)}\nonumber\\
&+\sum_{nb}\frac{\left(  -1\right)  ^{J+j_n-j_b}}{\left[  J\right]  }%
Z_{J}(bmna)Y_{n\left(  b\rightarrow\kappa_{n}\right)}\nonumber\\
&  +\sum_{nb}\frac{1}{\left[  J\right]  }Z_{J}(nmba)X_{n\left(  b\rightarrow
\kappa_{n}\right)  }\,.
\end{align}
\end{subequations}
It is worth remembering that the normalization condition in the $X$-$Y$ space differs from the conventional normalization. The former reads [see Eq.~\eqref{Eq:RPA-orthonormality}]
\begin{equation}
\sum_{ma}\left(\left\vert \chi_{ma}^{\mu}\right\vert ^{2}-\left\vert \eta_{ma}^{\mu
}\right\vert ^{2}\right)=\mathrm{sign}\left(  \omega^{\mu}\right)\,,
\end{equation}
which translates into
\begin{equation}
\sum_{ma}\left(\left\vert X_{m\left(  a\rightarrow\kappa_{m}\right)  }^{\mu
}\right\vert ^{2}-\left\vert Y_{m\left(  a\rightarrow\kappa_{m}\right)  }%
^{\mu}\right\vert ^{2}\right)=\mathrm{sign}\left(  \omega^{\mu}\right)  .
\end{equation}
Additionally, the symmetry property of the eigensystem now reads as follows:  for every pair
$\left\{  \omega_{\mu},\left(  X_{m\left(  a\rightarrow\kappa_{m}\right)
}^{\mu},Y_{m\left(  a\rightarrow\kappa_{m}\right)  }^{\mu}\right)  \right\}  $
there is a negative eigenfrequency counterpart $\left\{  -\omega_{\mu
},\left(  Y_{m\left(  a\rightarrow\kappa_{m}\right)  }^{\mu},X_{m\left(
a\rightarrow\kappa_{m}\right)  }^{\mu}\right)  \right\}  .$

Furthermore, using the Wigner-Eckart theorem, we arrive at the RPA-dressed reduced matrix elements,  
\begin{subequations}
    \begin{align}
\langle m||Z^{\mathrm{RPA}}||a\rangle &  =\left(  \varepsilon_{a}%
-\varepsilon_{m}+\omega\right)  \nonumber\\
&\times\sum_{\mu}\frac{\mathrm{sign}\left(
\omega^{\mu}\right)  \,R_{\mu}}{\omega-\omega_{\mu}}X_{m\left(  b\rightarrow
\kappa_{n}\right)  }^{\mu} \,,\\
\langle a||Z^{\mathrm{RPA}}||m\rangle &  =\left(  \varepsilon_{a}%
-\varepsilon_{m}-\omega\right)  \left(  -1\right)  ^{m-a+J}\nonumber\\
&\times\sum_{\mu}%
\frac{\mathrm{sign}\left(  \omega^{\mu}\right)  \,R_{\mu}}{\omega-\omega_{\mu
}}Y_{m\left(  b\rightarrow\kappa_{n}\right)  }^{\mu} \,,\label{Eq_RPApm}
\end{align}
\end{subequations}
where  the ``residuals'' $R_{\mu}$ are defined as
\begin{align}
  R_{\mu}   &\equiv\sum_{nb}\left(  X_{n\left(  b\rightarrow\kappa_{n}\right)
}^{\mu}\langle n||z||b\rangle\right.\nonumber\\
&+\left.\left(  -1\right)  ^{j_n-j_b+J}Y_{n\left(
b\rightarrow\kappa_{n}\right)  }^{\mu}\langle b||z||n\rangle\right)\,.
\end{align}

This concludes the derivation of our method. Some further simplifications are possible,
like reduction to positive frequency summations and we leave these straightforward steps to the reader. Beyond offering a one-shot solution to the RPA equations,
our approach is beneficial in evaluating  matrix elements for multiple transitions. Indeed, if one stores the eigenvectors $\left(  X_{n\left(  b\rightarrow\kappa
_{n}\right)  }^{\mu},Y_{n\left(  b\rightarrow\kappa_{n}\right)  }^{\mu
}\right)  $, the eigenvalues $\omega_{\mu}$, and the residuals $R_{\mu}$, then the
dressed matrix elements are easily assembled for any given driving frequency $\omega$. 

Finally, a numerical evaluation of the derived expressions requires single-particle orbital basis sets. In the main text, we use both the DHF and BO finite basis sets. These were described in Appendices~\ref{App:Finite-Basis-Set} and \ref{App:BO}, respectively.

\bibliographystyle{apsrev4-2}
\bibliography{bao}
\end{document}